\documentstyle[12pt,epsf,aaspp4]{article}

\input psfig.sty
\def\msun{{\rm\,M_\odot}}

\newcommand{\etal}{et al.\ }

\newcommand{\lya}{Ly$\alpha$ }

\def\etal   {{et~al.}\ }

\def\zsun{{\rm\,Z_\odot}}
\def\msun{{\rm\,M_\odot}}

\def\vol#1  {{{#1}{\rm,}\ }}

\def\etal{et al.\ }

\def\cf{{cf.}\ }

\begin{document}
\title{Signatures of Galactic Superwinds: Inhomogeneous Metal Enrichment of
the Lyman Alpha Forest} \author{Renyue Cen\altaffilmark{1}, Kentaro Nagamine
\altaffilmark{2} and  Jeremiah P. Ostriker\altaffilmark{3}}
\altaffiltext{1} {Princeton University Observatory, Princeton University,
Princeton, NJ 08544; cen@astro.princeton.edu}
\altaffiltext{2}{Harvard-Smithsonian Center for Astrophysics,
60 Garden Street, Cambridge, MA 02138; knagamin@cfa.harvard.edu}
\altaffiltext{3} {Princeton University Observatory, Princeton University,
Princeton, NJ 08544; jpo@astro.princeton.edu}

\begin{abstract}

We investigate possible signatures of feedback from
galactic superwinds on the metallicity of the \lya forest,
using a set of high resolution hydrodynamic simulations of
a $\Lambda$ cold dark matter model.
Simulations produce metals self-consistently, based on
one single parameter, the metal yield, which in turn is
constrained by metallicity in the intra-cluster gas.
We follow metals as a separate density species.
For the metallicity of \lya clouds with column density
of $N_{HI}\sim 10^{14.5}-10^{15.5}$ cm$^{-2}$ at $z=2-4$
we find reasonable agreement between
simulations, both with and without GSW, and observations (Schaye \etal). A
unique signature and sensitive test of GSW is, however,
provided by lower density regions
with gas density of $\rho/\langle\rho\rangle = 0.01-1.0$ and a corresponding
column density of $10^{12}-10^{14}$ cm$^{-2}$.
Without GSW we predict that both the mean and median
metallicity of \lya clouds in this column density range at $z=2-4$ should
have $Z\le 10^{-3}\zsun$,
since these small systems support little star formation.
GSW contaminate these regions, however, and also
there is a significant fraction ($\sim 25\%$)
of \lya clouds in this column density range which have a high metallicity
excess of $10^{-2}\zsun$, resulting in a mean metallicity
of $\sim 10^{-2}\zsun$.
In addition, we find that there is a minimum in the median metallicity for
clouds of $N_{HI}\sim 10^{13}-10^{14}$ cm$^{-2}$ in the case
with GSW,
whereas without GSW the metallicity decreases monotonically and rapidly with
decreasing column density.
Finally, we predict that the ratio of secondary (e.g., N) to primary metals
(e.g., O,C) is expected to be smaller by a factor of $10$
in clouds of $N_{HI}\sim 10^{14.5}$ cm$^{-2}$ compared to that
in large galaxies; this factor increases to $\ge 50$ for
$N_{HI}\le 10^{13.5}$ cm$^{-2}$.
\end{abstract}


\section{Introduction}

Direct observational evidence for feedback from galactic superwinds (GSW)
originating in starburst galaxies is ubiquitous both at
low redshift
(e.g., McCarthy, Heckman, \& van Breugel 1987;
Heckman, Armus, \& Miley 1987;
Papaderos \etal 1994;
Marlowe \etal 1995;
Lehnert \& Heckman 1996;
Bland \& Tully 1988;
Filippenko \& Sargent 1992;
Dahlem, Weaver, \& Heckman 1998;
Heckman \etal 1998;
Martin 1999;
Yoshida, Taniguchi, \& Murayama 1999;
Veilleux \etal 1999;
Della Ceca \etal 1999;
Veilleux, Shopbell, \& Miller 2001;
Rupke, Veilleux, \& Sanders 2002;
Martin, Kobulnicky, \& Heckman 2002)
and at high redshift (e.g., Franx \etal 1997;
Pettini \etal 1998;
Dawson \etal 2002;
Pettini \etal 2001, 2002;
Adelberger 2003; 
Adelberger \etal 2003).
In addition, two lines of indirect but independent observational evidence
point to the existence or need of GSW.
First, the low-to-moderate density regions
of the intergalactic medium (IGM)
in \lya clouds at $z\sim 2-3$
have already been enriched with
metals to significant levels
(e.g., Tytler \etal 1995; Songaila \& Cowie 1996;
Bergeron \etal 2002)
to a level that would be difficult to achieve
by sources embedded in those regions.
And in the warm-hot intergalactic medium
at low redshift (e.g., Tripp, Savage, \& Jenkins 2000;
Fang \etal 2002; Nicastro \etal 2002;
Mathur, Weinberg, \& Chen 2003)
metals are seen that presumably originate from the galaxies
central to these regions.
Second, a substantial,
non-gravitational heating source of the intra-cluster medium
may be needed to produce the observed
X-ray cluster luminosity-temperature relation
and its evolution (e.g., Kaiser 1991;
White 1991; David, Forman, \& Jones 1991;
Metzler \& Evrard 1994; Navarro, Frenk, \& White 1995;
Pen 1999; Ponman, Cannon, \& Navarro 1999;
Balogh, Babul, \& Patton 1999;
Loewenstein 2000; Wu, Fabian, \& Nulsen 2000;
Lloyd-Davies, Ponman, \& Cannon 2000;
Brighenti \& Mathews 2001; Neumann \& Arnaud 2001;
Borgani \etal 2001; Voit \& Bryan 2001;
Tozzi \& Norman 2001; Dav\'e \etal 2001;
Babul \etal 2002; Bialek, Evrard, \& Mohr 2001;
McCarthy, Babul, \& Balogh 2002;
Afshordi \& Cen 2002;
Voit \etal 2002).
GSW may play a significant role in the transport of metal-enriched matter to
lower density regions outside galaxies
and may help provide
the requisite non-gravitational heating source for the cluster gas.

In contrast to notable successes of cosmological hydrodynamic
simulations of the IGM that has not been
intimately involved in star formation, such as the Lyman-alpha forest (Cen
\etal 1994; Zhang \etal 1995; Hernquist \etal 1996;
Miralda-Escud\'e \etal 1996;
Bond \& Wadsley 1997;
Theuns \etal 1998),
few calculations have been made
to investigate the impact of GSW on the IGM
in a coherent fashion.
So far, most brute-force cosmological hydrodynamic simulations
do not include the feedback effects of the GSW.
For those simulations with GSW included
(Cen, \& Ostriker 1992, 1993a,b; Cen \etal 1994;
Gnedin \& Ostriker 1997; Gnedin 1998; Cen \& Ostriker 1999b;
Springel \& Hernquist 2003; Kay \etal 2002; Theuns \etal 2002),
the obtained results are often paradoxical;
for example, in very high resolution simulations 
dense regions tend to
radiate away most of the GSW feedback energy
and thus largely suppress its effect.
This is for the most part due to
the limited (and sometimes mismatched spatial and mass)
numerical resolutions
with the inability to properly represent, in cosmological simulations, a
multi-phase ISM,
where a GSW originates.
On the other hand, simulations with lower resolution, limited by available
computer power, but with a crude
multi-phase medium treatment (e.g., Cen \& Ostriker 1999a,b)
appear to be able to
exert substantial feedback energy on the general IGM surrounding
galaxies.
But the very limited resolution of these simulations
does not allow us to draw reliable quantitative conclusions.

Although it is generally accepted that
energy from collective supernova explosions and stellar winds
should be powering the GSW (Ostriker \& Cowie 1981;
see Aguirre 1999 for a role
that radiation pressure on dust plays in driving outflows),
the complex structure of the interstellar medium (ISM)
(McKee \& Ostriker 1977)
and the IGM
makes quantitative calculations of GSW and subsequent evolution
a daunting task,
which certainly requires treatment of a multi-phase
medium and may necessitate the explicit
inclusion of magnetic fields
(Koo \& McKee 1992a,b;
Smith 1996;
Suchkov \etal 1996;
Nath \& Trentham 1997;
Hartquist, Dyson, \& Williams 1997;
Gnedin \& Ostriker 1997;
Gnedin 1998;
Mac Low \& Ferrara 1999;
Cen \& Ostriker 1999b;
Ferrara, Pettini, \& Shchekinov 2000;
Madau, Ferrara, \& Rees 2001;
Aguirre \etal 2001;
Mori, Ferrara \& Madau 2002;
Scannapieco, Thacker, \& Davis 2001;
Scannapieco, Ferrara, \& Madau 2002;
Thacker, Scannapieco, \& Davis 2002;
White, Hernquist, \& Springel 2002;
Dyson, Arthur, \& Hartquist 2002;
Springel \& Hernquist 2003).
Significant progress has been made recently
to provide a better treatment of the multi-phase interstellar
medium (Yepes \etal 1997;
Elizondo \etal 1999a,b;
Hultman \& Pharasyn 1999;
Ritchie \& Thomas 2001;
Springel \& Hernquist 2003) but
the generation of GSW is far from being adequately modeled.
Clearly, a combination of both high resolution and
detailed multi-phase medium treatment
(including magnetic fields and cosmic rays)
is requisite before
our understanding of the interactions
between galaxy formation and IGM
can be considered to be truly satisfactory.

But we will follow a somewhat different approach.
We will not attempt to model the complex physics
which determines how much of the SN energy
produced within the galaxies can escape the galaxies.
Rather we will inject energy
directly into the medium surrounding the galaxies in
a fashion that drives GSW
and we will adjust the energy input to match the observed GSW.
Direct and empirical determination
of the output of energy and metal enriched gas
from GSW is, in principle, possible (Chevalier \& Clegg 1985),
although, in practice, a complete account of the energy and mass output
(especially the hot component at the X-ray band)
from GSW may require more involved work (Strickland \& Stevens 2000).
Nevertheless, direct observational
determinations of mass and energy loss rates from GSW
have yielded very interesting results
(see Heckman 2001 for a recent review),
with observations of both low redshift
starburst galaxies and high redshift Lyman Break Galaxies (LBGs)
indicating that both mass and energy outflows
from GSW are comparable to those supplied by the interior
starburst.

Thus, we accept at the outset our inability to correctly
model the detailed structure of the ISM within
these galaxies or the {\it generation} of the GSW.
Rather, we will simply  assume a proportionality
between the star formation rate
$\dot M_*$ in the system
and the output by that system
of wind mass flux and energy flux,
since there is a sound observational basis for this assumption.
So we will assume that the energy {\it output}
in a GSW is related to the star formation
rate by
$\dot E_{GSW} = e_{GSW} c^2 \dot M_*$
(where $c$ is the speed of light)
and the mass output is
$\dot M_{GSW} = \beta_{GSW} \dot M_*$.
The two adjustable parameters ($e_{GSW}$, $\beta_{GSW}$)
are then determined by a fit to observations
(e.g., Pettini \etal 2001,2002; Heckman 2001),
specifically, the two observed parameters -
the mass flow rate and the wind velocity,
and our subsequent computations are utilized to determine
the {\it effects} of the consequent GSW
on the metallicity distribution within
the IGM,
the shock-heating input to the IGM
and the modification/regulation
of subsequent galaxy formation.
In brief, we seek to model the consequences
not the causes of GSW
feedback, and this is something that we think
our codes are well designed to do.
It should be stated that our approach is clearly incapable
of fully solving the feedback process, since it is
a phenomenological approach.
But this is a major step forward to understand
the effects of GSW, given the current state of knowledge
which leaves the physics of generation of GSW largely unconstrained.

It may be useful to put this in a historical context.
A decade ago the focus of cosmological simulations (e.g., Cen \etal 1994)
was to fit the observed \lya forest into the picture of
modern hierarchical structure formation theory.
The result was the emergence of the now standard theory
for \lya forest based on the growth/collapse of small
scale density perturbations at moderate redshift.
In this post-{\it WMAP (Wilkinson Microwave Anisotropy Probe)} era,
research focus for \lya forest has become to provide answers to the
following question:
how does galaxy formation affect the properties of \lya forest
and how is the power spectrum of primordial density fluctuation 
on small-scales reconstructed from \lya forest flux distribution
subject to various processes related to galaxy formation?
This paper attempts to provide some partial answers to
the first half of the question.
In this first of a series of papers focusing on the
effects of GSW on the IGM and subsequent galaxy/star formation,
we will investigate the effect of GSW on the metal enrichment
of the \lya forest at high redshift ($z=2-4$),
which contains most of the mass as well as volume of the IGM then. The
outline of this paper is as follows.
The simulation details are given in \S 2.
In \S 3 we give detailed results
and we conclude in \S 4.

\section{Simulations}

Numerical methods of the cosmological hydrodynamic code
and input physical ingredients
have been described in detail in an earlier paper
(Cen \etal 2003).
We will name the code TIGER,
{\bf T}vd for {\bf I}ntergalactic medium and {\bf G}alaxy 
{\bf E}volution and fo{\bf R}mation.
Briefly, the simulation integrates five sets of equations
simultaneously:
the Euler equations for gas dynamics,
rate equations for different hydrogen and helium
species at different ionization states,
the Newton's equations of motion for dynamics of collisionless particles,
the Poisson's equation for obtaining the gravitational potential field and
the equation governing the evolution of the intergalactic
ionizing radiation field,
all in cosmological comoving coordinates.
Note that the cosmological (frequency dependent)
radiation field is solved for self-consistently, rather than
being a separate input to the modeling.
The gasdynamical equations are solved using
the TVD (Total Variation Diminishing) shock capturing code 
(Ryu \etal 1993) on an uniform mesh.
The rate equations are treated using sub-cycles within a hydrodynamic time
step due to much shorter ionization time-scales
(i.e., the rate equations are very ``stiff").
Dark matter particles are advanced in time using the standard
particle-mesh (PM) scheme.
The gravitational potential on an uniform mesh is solved
using the Fast Fourier Transform (FFT) method.

The initial conditions adopted are those
for Gaussian processes with the phases
of the different waves being random and uncorrelated.
The initial condition is generated by the
COSMICS software package kindly provided by E. Bertschinger (2001).

Cooling and heating
processes due to all the principal line and continuum atomic
processes for a plasma of primordial composition with
additional metals ejected from star formation (see below),
Compton cooling due to the microwave background
radiation field and Compton cooling/heating due to
the X-ray and high energy background
are computed  in a time-dependent, non-equilibrium fashion.
The cooling due to metals is computed
using a code based on the Raymond-Smith code (Raymond, Cox, \& Smith 1976)
assuming ionization
equilibrium (Cen \etal 1995).

We follow star formation using
a well defined  
prescription
used by us in our earlier work
(Cen \& Ostriker 1992,1993) and
similar to that of other investigators
(Katz, Hernquist, \& Weinberg 1992;
Katz, Weinberg, \& Hernquist 1996;
Steinmetz 1996;
Gnedin \& Ostriker 1997).
A stellar particle of mass
$m_{*}=c_{*} m_{\rm gas} \Delta t/t_{*}$ is created
(the same amount is removed from the gas mass in the cell),
if the gas in a cell at any time meets
the following three conditions simultaneously:
(i) contracting flow, (ii) cooling time less than dynamic time, and  (iii)
Jeans unstable,
where $\Delta t$ is the time step, $t_{*}={\rm max}(t_{\rm dyn}, 10^7$yrs),
$t_{dyn}=\sqrt{3\pi/(32G\rho_{tot})}$ is the dynamical time of the cell,
$m_{\rm gas}$ is the baryonic gas mass in the cell and
$c_*=0.07$ is star formation efficiency.
Each stellar particle has a number of other attributes at birth, including
formation time $t_i$, initial gas metallicity
and the free-fall time in the birth cell $t_{dyn}$.
The typical mass of a stellar particle in the simulation
is about one million solar masses;
in other words, these stellar particles are like
coeval globular clusters.

Stellar particles are subsequently treated dynamically
as collisionless particles.
But feedback from star formation is allowed in three forms:
ionizing UV photons, supernova kinetic energy (i.e., GSW), and metal-enrich
gas, all being proportional to the local star formation rate.
The temporal release of all three feedback components at time $t$
has the same form:
$f(t,t_i,t_{dyn}) \equiv (1/ t_{dyn})
[(t-t_i)/t_{dyn}]\exp[-(t-t_i)/t_{dyn}]$. Within a time step $dt$, the
released GSW energy to the IGM, ejected mass from stars into the IGM and
escape UV radiation energy are
$e_{GSW} f(t,t_i,t_{dyn}) m_* c^2 dt$,
$e_{mass} f(t,t_i,t_{dyn}) m_* dt$
and
$f_{esc}(Z) e_{UV}(Z) f(t,t_i,t_{dyn}) m_* c^2 dt$.
We use the Bruzual-Charlot population synthesis code 
(Bruzual \& Charlot 1993; Bruzual 2000)  
to compute the intrinsic metallicity-dependent
UV spectra from stars with Salpeter IMF (with a lower and upper mass 
cutoff of $0.1\msun$ and $125\msun$).
Note that $e_{UV}$ is no longer just a simple coefficient but
a function of metallicity.
The Bruzual-Charlot code gives
$e_{UV}=(1.2\times 10^{-4}, 9.7\times 10^{-5}, 8.2\times 10^{-5}, 7.0\times 10^{-5}, 5.6\times 10^{-5}, 3.9\times 10^{-5} ,1.6\times 10^{-6})$ at
$Z/\zsun=(5.0\times 10^{-3}, 2.0\times 10^{-2}, 2.0\times 10^{-1}, 4.0\times 10^{-1}, 1.0, 2.5 ,5.0)$. 
We also implement a gas metallicity dependent ionizing photon
escape fraction from galaxies in the sense that higher metallicity  hence
higher dust content galaxies are assumed to allow a lower
escape fraction; we adopt the escape fractions of
$f_{esc}=2\%$ and $5\%$ (Hurwitz \etal 1997; Deharveng \etal 2001;
Heckman \etal 2001)
for solar and one tenth of solar metallicity, respectively,
and interpolate/extrapolate using a linear log form of metallicity. In
addition, we include the emission from
quasars using the spectral form observationally
derived by Sazonov, Ostriker, \& Sunyaev (2004),
with a radiative efficiency in terms of stellar mass
of $e_{QSO}=2.5\times 10^{-5}$ for $h\nu>13.6$eV.
Finally, hot, shocked regions (like clusters of galaxies)
emit ionizing photons due to bremsstrahlung radiation,
which are also included.
The UV component is simply averaged over the box,
since the light propagation time across our box
is small compared to the time steps.
The radiation field (from $1$eV to $100$keV)
is followed in detail with allowance for
self-consistently produced radiation sources and sinks in the simulation box
and for cosmological effects, i.e., radiation transfer
for the mean field $J_\nu$ is computed with stellar,
quasar and bremsstrahlung sources and sinks due to \lya clouds etc. In
addition, a local optical depth approximation is adopted to crudely mimic
the local shielding effects: each cubic cell is flagged with six
hydrogen ``optical depths" on the six faces, each equal to the product of
neutral hydrogen density, hydrogen ionization cross section and scale
height,
and the appropriate mean from the six values is then calculated;
equivalent ones for neutral helium and singly-ionized helium are also
computed. In computing the global sink terms for the radiation field
the contribution of each cell is subject to the shielding
due to its own ``optical depth".
In addition, in computing the local ionization and cooling/heating balance
for each cell the same shielding is taken into account
to attenuate the external ionizing radiation field.

GSW energy and ejected metals
are distributed into 27 local gas cells centered at the stellar particle in
question, weighted by the specific volume of each cell.
We fix $e_{mass}=0.25$.
GSW energy injected into the IGM is included with an adjustable
efficiency (in terms of rest-mass energy of total formed stars) of
$e_{GSW}$, which is normalized to observations for our fiducial simulation
with $e_{GSW}=3\times 10^{-6}$.
If the ejected mass and associated energy propagate into a vacuum, the
resulting velocity of the ejecta would be
$(2e_{GSW}/e_{mass})^{1/2}c=1469$km/s.
After the ejecta has accumulated an amount of mass comparable to its initial
mass, the velocity may slow down to a few hundred km/s.
We assume this velocity would roughly correspond to the observed
outflow velocities of LBGs (e.g., Pettini \etal 2002).
We also make simulations with no GSW and with stronger GSW
to investigate the effects of GSW on IGM.

We do not separately make any adjustments to fit to the observed
distributions and evolution of metals,
but assume a specific efficiency of metal formation,
an ``yield" (Arnett 1996), $y_0=0.02$,
the percentage of stellar mass that is ejected back into IGM as metals.
We note that $y_0=Z_{ejecta} e_{mass}$;
since $y_0=0.02$ and $e_{mass}=0.25$,
it implies that the ejecta metallicity is
$Z_{ejecta}=0.08=4\zsun$.
Metals in the IGM (assuming the standard solar composition)
are followed as a separate variable
(analogous to the total gas density) with the same hydrocode.
In addition, we implement another density variable to
keep track of the reprocessed, i.e., secondary metals in the ejecta,
which is proportional to the metallicity of the gas from 
which the star was formed.

Since we are interested in the metallicity of the IGM,
it is legitimate to question whether our adopted constant
metal yield is reasonable.
We can not answer this question  from first principles.
Rather, we will consider a physically motivated case,
where the metallicity yield is a function of gas metallicity
out of which stars are formed.
It is thought that the initial mass function of stars
formed out of low metallicity gas may contain relatively
more high mass stars thus produce a higher yield
(for references to the original literature see 
Ricotti \& Ostriker 2004).
We adopt this view and consider a scenario with varying yield
by making a correction to the computed yield as described below.
We present results for both the case of constant yield and
metallicity dependent yield to indicate
the uncertainties and/or adjustability of the results.

Let the yield be $y(Z)$.
It can then be shown that the final corrected metallicity of a region with
computed metallicity $Z_c$ will be
$Z_v=Z_c y(Z_c)/(y_0 f(Z_c))$, where
$f(Z_c) = (y(Z_c)/Z_c)\int_0^{Z_c}dx/y(x)$.
We somewhat arbitrarily set the form of $y(Z)$ to be
$y(Z)=(5.0+45.0(1-\exp(-Z/B))^{-1}$, which gives
$y=0.02$ for $Z\gg B$ and $y=0.2$ for $Z\ll B$.
Thus we have adopted a higher yield of $0.2$ for metal-free stars, which may
be a reasonable choice if IMF of metal-free stars are top-heavy (Woosley \&
Weaver 1995).
The transition metallicity $B$ is uncertain
but we use $10^{-3}\zsun$ for the illustration (Bromm \& Loeb 2003; Fang \&
Cen 2004).

The results reported on here are based on new simulations of
a {\it WMAP}-normalized (Spergel \etal 2003) cold dark matter model with 
a cosmological constant:
$\Omega_M=0.29$, $\Omega_b=0.047$, $\Omega_{\Lambda}=0.71$, $\sigma_8=0.85$,
$H_0=100 h {\rm km s}^{-1} {\rm Mpc}^{-1} = 70 {\rm km} s^{-1} {\rm Mpc}^{-1}$ 
and $n=1.0$.
Seven simulations with varying box size, resolution
and input physics are made, as listed in Table 1.
Mass resolution is extremely important for an analysis of this type as it is
the lowest mass systems in relatively low density
regions that contain the stars which are most suitable for
contamination of the low and moderate density IGM.
The mass resolutions are considerably better than those
of most cosmological
simulations, but the spatial resolution, while significantly inferior to
that obtained in both the SPH and AMR schemes, is, we believe,
adequate for the present purpose.
The coarser spatial resolution among the listed
simulations is smaller than the Jeans length
of photoionized IGM at $z=2-4$ by
a factor greater than $10$ but only marginally resolve
some small galaxies of total mass $10^9\msun$.
But our higher resolution simulation indicates
that results are not significantly affected.

\begin{deluxetable}{ccccccc}
\tablecolumns{8}
\tablewidth{0pc}
\tablecaption{A List of Simulations}
\tablehead{
\colhead{Run} & \colhead{Label} & \colhead{Box}
& \colhead{Spatial Res}  & \colhead{Mass Res}
& \colhead{$k_{max}$} & \colhead{$e_{GSW}$} }
\startdata
1 & N432L11M   & 11$h^{-1}$Mpc & 25$h^{-1}$kpc & $2.1\times 10^5\msun$ &
$123h{\rm Mpc}^{-1}$ & $3\times 10^{-6}$   \\ 
2 & N432L11L   & 11$h^{-1}$Mpc & 25$h^{-1}$kpc & $2.1\times 10^5\msun$ & 
$123h{\rm Mpc}^{-1}$ & $0$                 \\
3 & N432L11H   & 11$h^{-1}$Mpc & 25$h^{-1}$kpc & $2.1\times 10^5\msun$ &
$123h{\rm Mpc}^{-1}$ & $1.5\times 10^{-5}$ \\ 
4 & N864L11M   & 11$h^{-1}$Mpc & 13$h^{-1}$kpc & $2.7\times 10^4\msun$ & 
$123h{\rm Mpc}^{-1}$ & $3\times 10^{-6}$   \\
5 & N864L22M   & 22$h^{-1}$Mpc & 25$h^{-1}$kpc & $2.1\times 10^5\msun$ &
$123h{\rm Mpc}^{-1}$ & $3\times 10^{-6}$   \\ 
6 & N432L11M8  & 11$h^{-1}$Mpc & 25$h^{-1}$kpc & $2.1\times 10^5\msun$ & 
$31h{\rm Mpc}^{-1}$  & $3\times 10^{-6}$   \\
7 & N432L11M32 & 11$h^{-1}$Mpc & 25$h^{-1}$kpc & $2.1\times 10^5\msun$ &
$7.7h{\rm Mpc}^{-1}$ & $3\times 10^{-6}$   \\ \hline
\enddata
\tablecomments{
The first and second columns give a numeric number for each run
and a label indicating the number of cells used (432 for $432^3$ cells and
864 for $864^3$ cells),
the box size (11$h^{-1}$Mpc and 22 $h^{-1}$Mpc)
and the level of GSW (`L' for low, `M' for median, and `H' for high).
The simulations labeled with $N432$ have $216^3$ dark matter particles,
whereas those labeled with $N864$ have $432^3$ dark matter particles. Box
sizes (third column)
and spatial resolution (fourth column)
are both in comoving units.
The fifth column is the
mean baryonic cell mass;
the corresponding dark matter particle mass
is $1.0\times 10^7\msun$ for Runs (1,2,3,5,6,7)
and $1.3\times 10^6\msun$ for Run 4.
The initial maximum wavenumber $k_{max}$ (sixth column)
for the input power spectrum is in comoving $h{\rm Mpc}^{-1}$.
The last column indicates the GSW strength.
}
\end{deluxetable}

The first simulation (N432L11M) is our fiducial one with
a GSW feedback that is approximately matched to observations
of Lyman break galaxies.
The second simulation (N432L11L) has negligible GSW,
while the third simulation (N432L11H) has GSW energy that
is higher than the fiducial run by a factor of $5$.
The higher resolution run (N864L11M) has twice as high
spatial resolution but with all other physics fixed the same and
is made to check the dependence of the results on the resolution
and the convergence of the results on resolution.
As will be shown below, a proper convergence has been
achieved for the problem in hand.
The larger simulation (N864L22M) is made to check the
dependence of the results on the box size and, as one would
have expected, the $11h^{-1}$Mpc box seems adequate
for the objects under investigation at high redshift.
The last two simulations (N432L11M8 and N432L11M32)
have the exact same input physics and resolutions as
the fiducial run but with the initial power spectrum
cutoff at $8$ and $32$ cells, respectively,
instead of $2$ cells in the fiducial run.
These two runs were made with the purpose of
isolating some of the effects due to small galaxies
forming from density fluctuations of high wave numbers.

The program used to generate synthetic \lya forest lines
here is the same one used in our previous papers
(Cen \etal 1994; Miralda-Escud\'e \etal 1996).
The only addition is that we have in addition the metal density,
which allows detailed computations of metallicity distributions
in the \lya forest.
All transmitted flux is computed with a FWHM of $6.6 {\rm km s}^{-1}$,
a sample pixel size of $2 {\rm km s}^{-1}$ and Gaussian noise added to
each pixel with a signal-to-noise ratio of $150$.
Mean decrement is chosen to match $\bar D=0.34$ (e.g., Press, Rybicki, \&
Schneider 1993) in all simulated spectra by adjusting
the background radiation field to facilitate a meaningful
comparison.

\section{Results}

Figure 1 shows the distributions of IGM temperature
for a typical slice of the indicated size.
The left panel is with GSW (Run 1: N432L11M) and
the right panel without GSW (Run 2: N432L11L).
It is clear that GSW do blow bubbles of hot gas,
which occupy a radius typically of hundreds of kpc.

Before proceeding to compute the metallicity distribution
in the \lya forest,
it is pertinent to ask if a substantial
GSW feedback on the IGM may spoil the excellent agreement
found between the predictions of the cold dark matter model
and the observed \lya forest
(Cen \etal 1994; Zhang \etal 1995; Hernquist \etal 1996;
Miralda-Escud\'e \etal 1996;
Rauch \etal 1997;
Croft \etal 1999;
McDonald \etal 2000).
The recent work of Theuns \etal (2002)
has clearly demonstrated that GSW mainly propagate
in the directions of lowest column density,
and filaments (producing most of the \lya forest)
are not significantly affected by GSW.
They show quantitatively that \lya forest statistics
such as column density distribution and
Doppler width distribution remain little changed
and the good agreement between cold dark matter model
and observations is, to the zero-th order,
unaltered by GSW to the concerned accuracies.
Here we will confirm their conclusions.
Figure 2 visually presents this point, showing
little alteration of the density distributions in filaments.
A joint examination of Figures 1,2 indicates
that GSW prefer to travel in the directions roughly
perpendicular to the filaments, as found by Theuns \etal (2002).

\begin{figure*}[t!]
\centering
\begin{picture}(500,200)
\psfig{figure=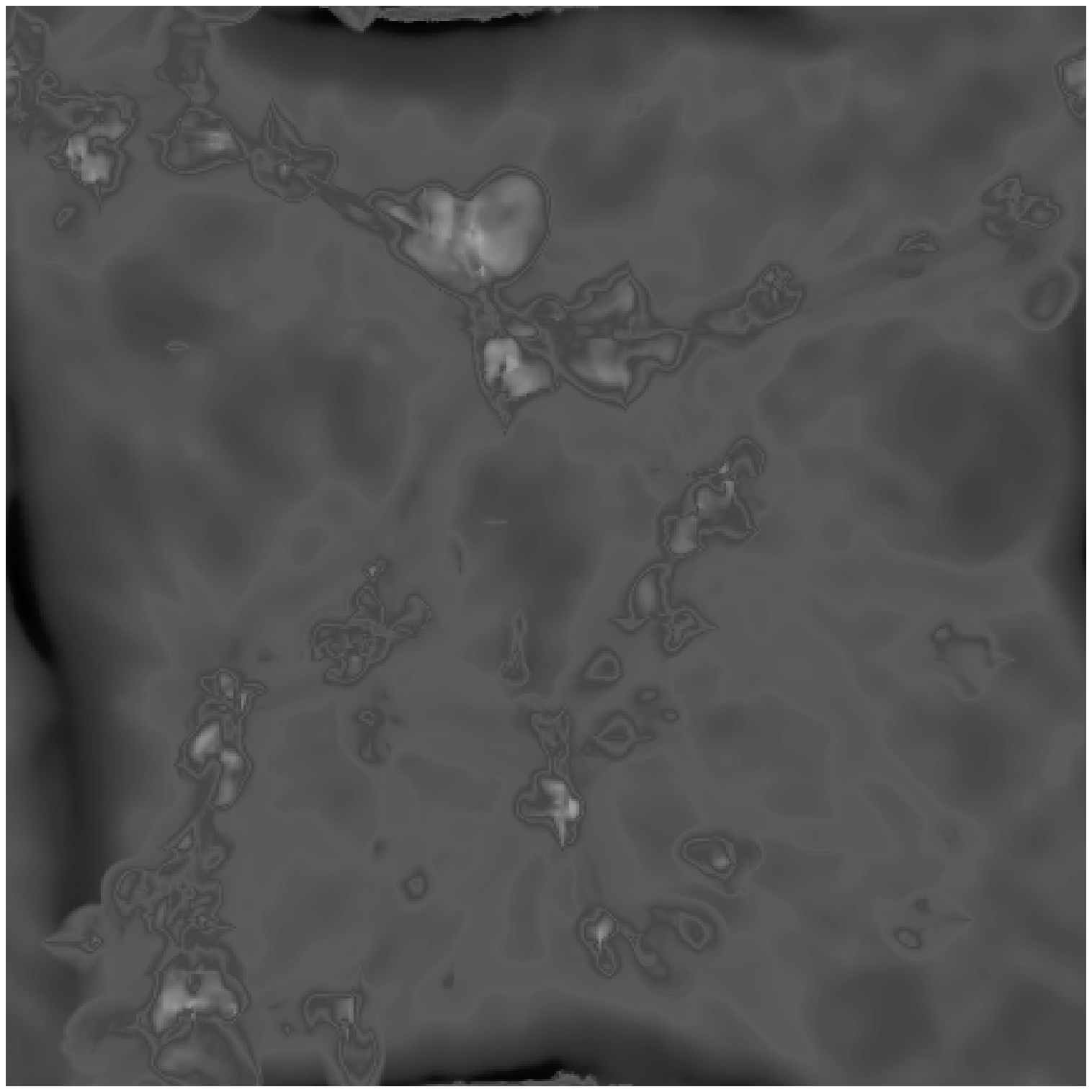,height=7.0cm,width=7.0cm,angle=0.0}
\end{picture}
\vskip -0.5cm
\centering
\begin{picture}(-50,-1000)
\psfig{figure=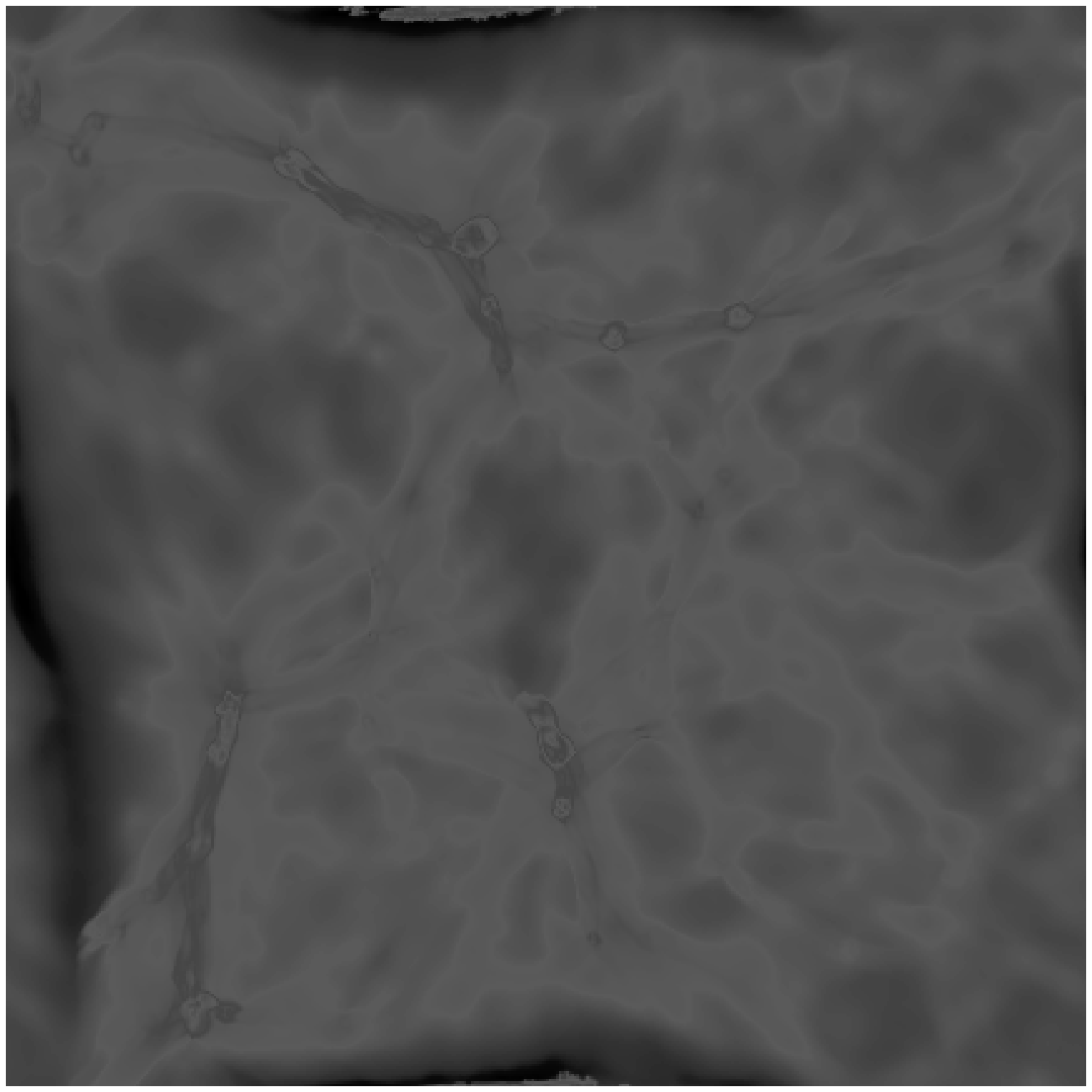,height=7.0cm,width=7.0cm,angle=0.0}
\end{picture}
\vskip -0.3cm
\centering
\begin{picture}(63,-1000)
\psfig{figure=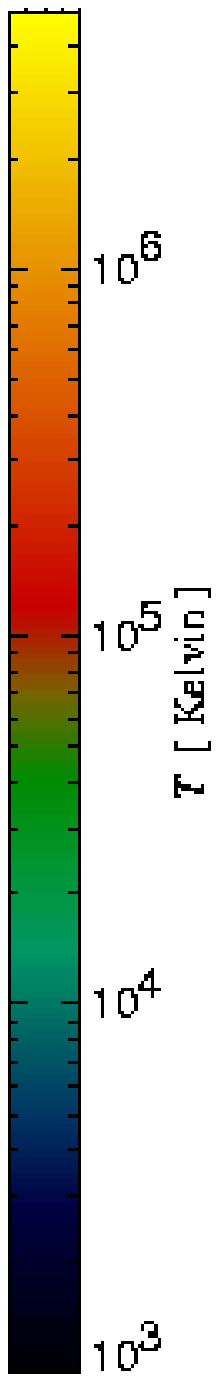,height=7.5cm,width=1.5cm,angle=0.0}
\end{picture}
\caption{
Projected temperature of a slice of size $11\times 11h^{-2}$Mpc$^2$ comoving
and a depth of $2.75 h^{-1}$Mpc comoving at redshift $z=3$
with (left panel) and without (right) GSW, respectively.
The strength of the GSW is normalized to LBG observations.
\label{fig1}}
\end{figure*}

\begin{figure*}[b!]
\centering
\begin{picture}(500,200)
\psfig{figure=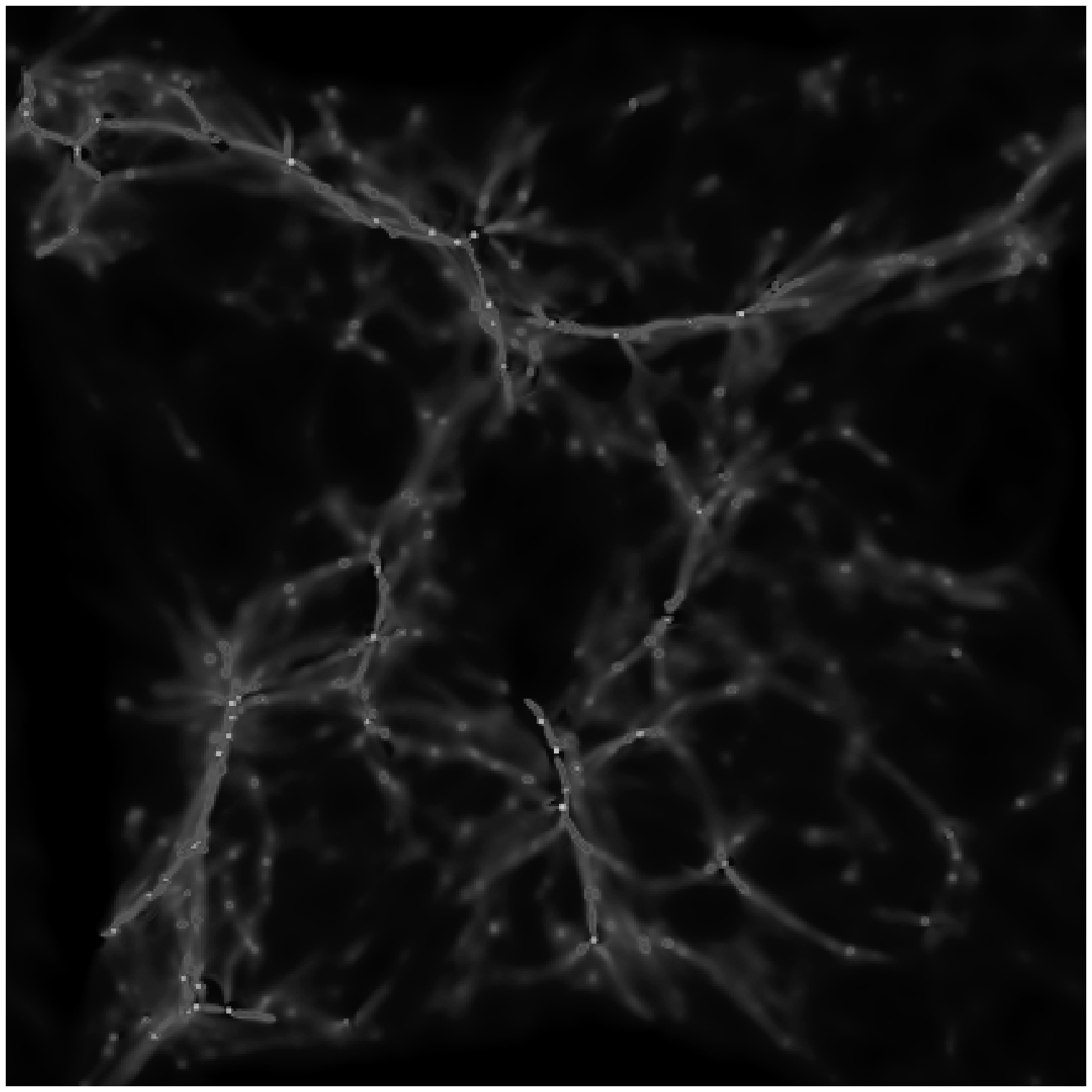,height=7.0cm,width=7.0cm,angle=0.0}
\end{picture}
\vskip -0.5cm
\centering
\begin{picture}(-50,-1000)
\psfig{figure=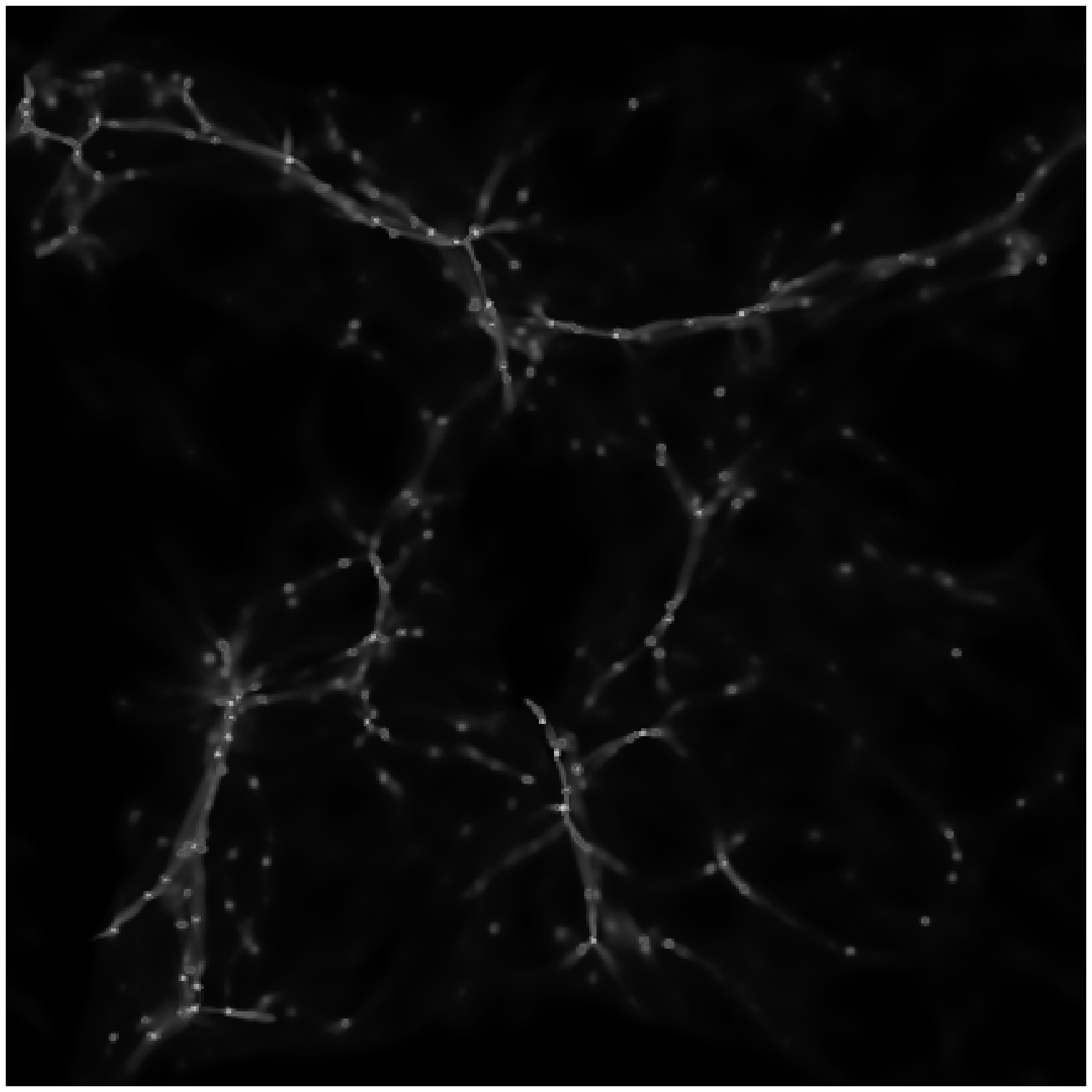,height=7.0cm,width=7.0cm,angle=0.0}
\end{picture}
\vskip -0.3cm
\centering
\begin{picture}(43,-3000)
\psfig{figure=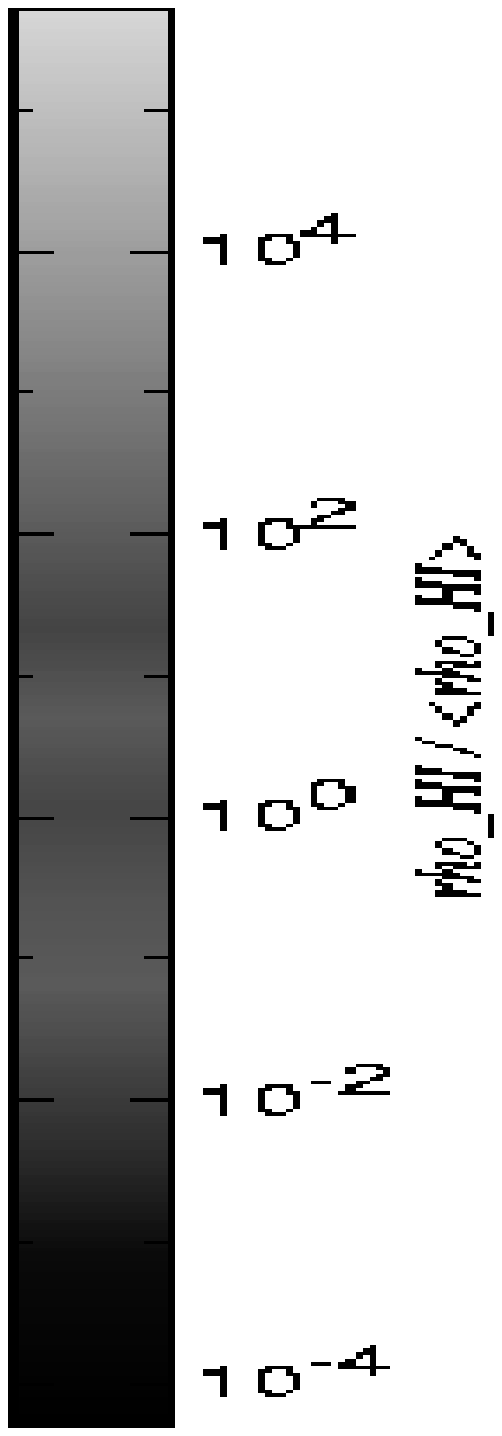,height=7.5cm,width=1.5cm,angle=0.0}
\end{picture}
\caption{
Projected neutral hydrogen overdensity of
the same slice as in Figure 1
with (left panel) and without (right) GSW, respectively.
\label{fig2}}
\end{figure*}

\begin{figure*}[t!]
\centering
\vskip 0.7cm
\begin{picture}(250,200)
\psfig{figure=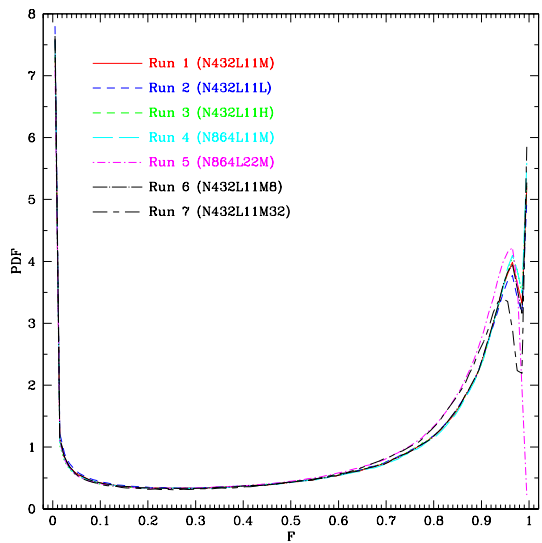,height=8.0cm,width=8.0cm,angle=0.0}
\end{picture}
\vskip -0.3cm
\centering
\caption{
shows the flux probability distribution function (PDF)
for the seven runs at $z=3$.
\label{fig3}}
\end{figure*}

\begin{figure*}[b!]
\centering
\begin{picture}(500,200)
\psfig{figure=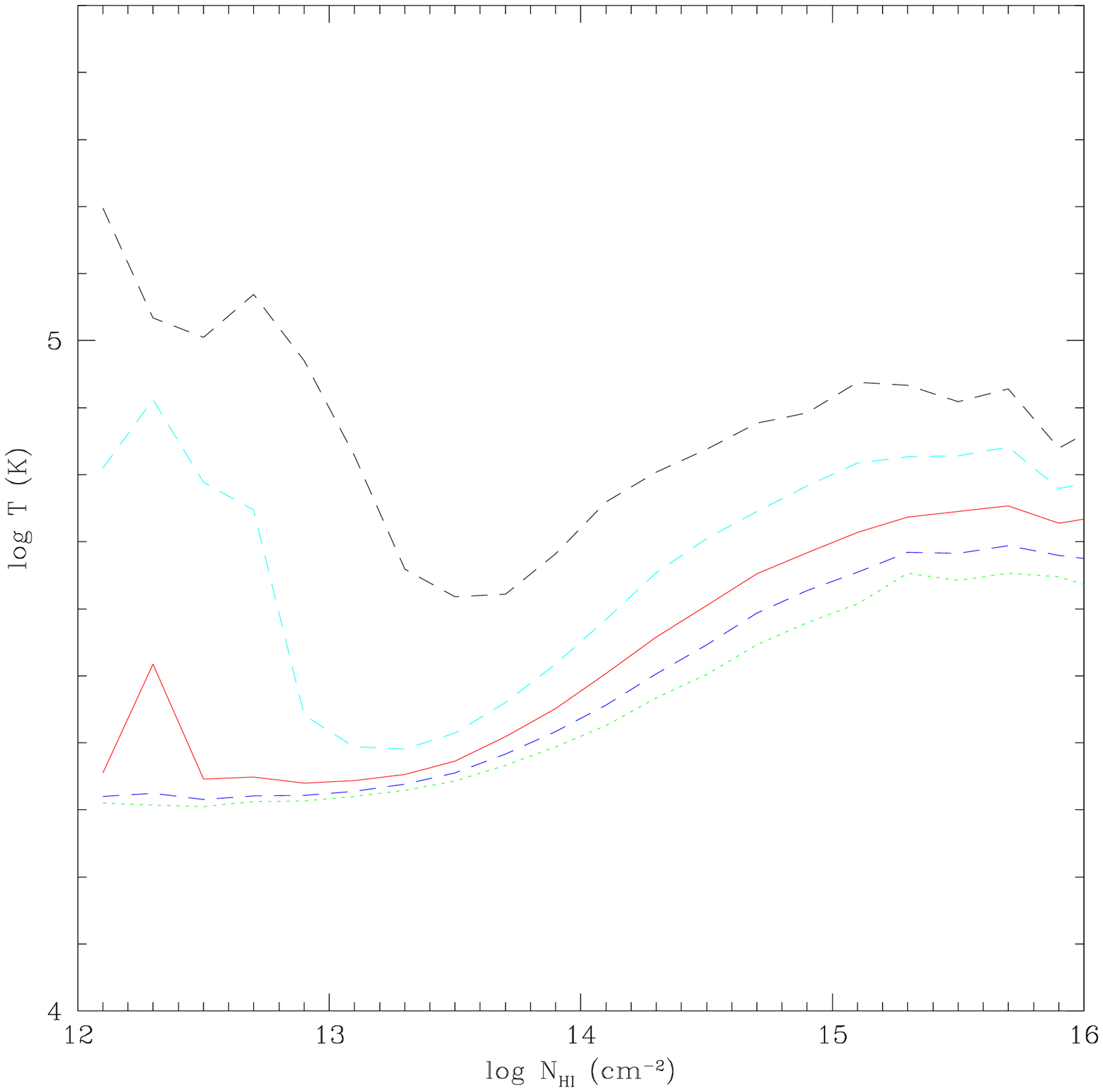,height=7.0cm,width=7.0cm,angle=0.0}
\end{picture}
\vskip -0.5cm
\centering
\begin{picture}(-50,-1000)
\psfig{figure=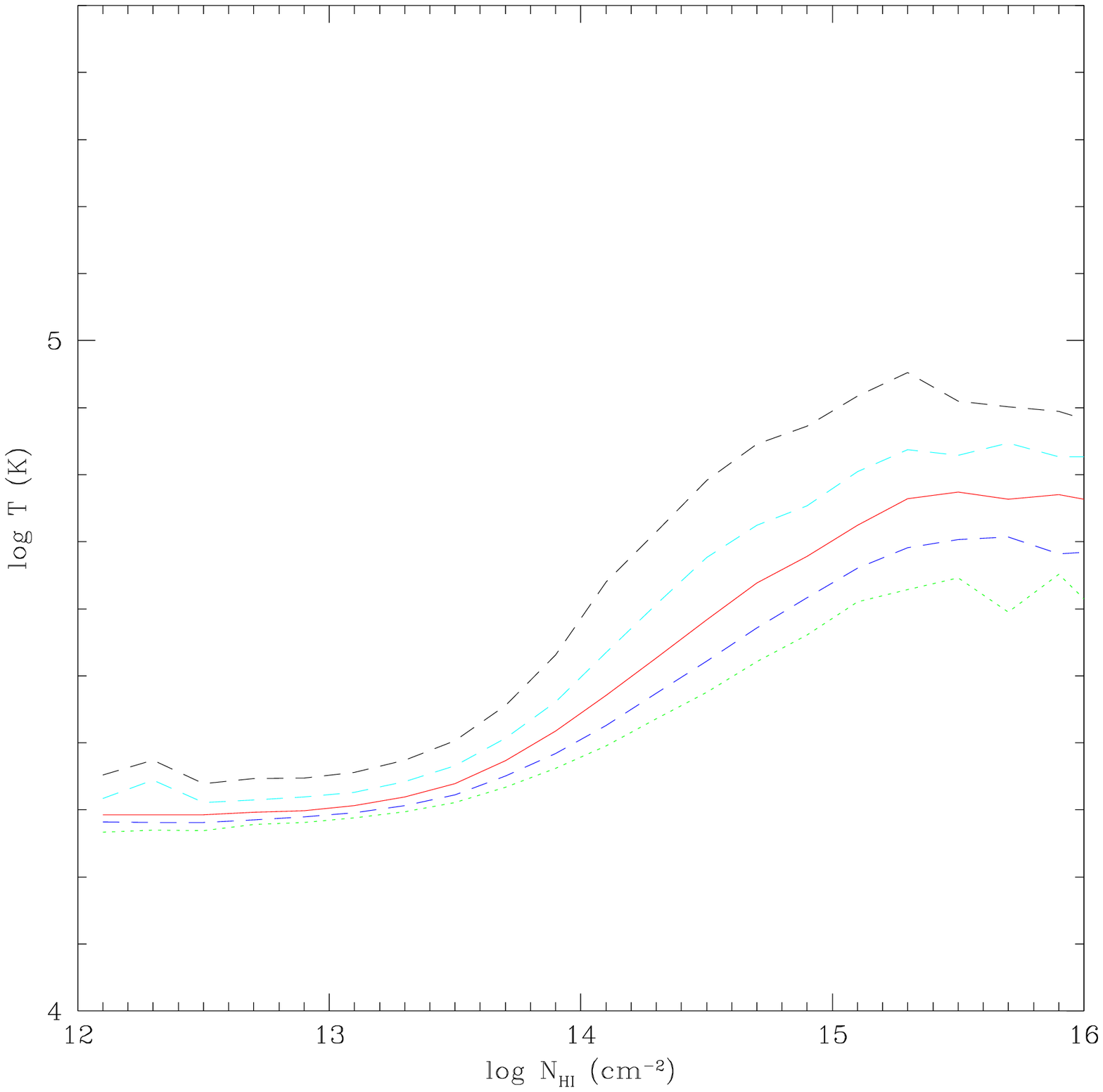,height=7.0cm,width=7.0cm,angle=0.0}
\end{picture}
\vskip -0.3cm
\centering
\caption{
shows the temperature as a function of \lya cloud
column density
for
the two cases with
(left panel; Run 1:N432L11M) and without (right panel; Run 2:N432L11L) GSW,
respectively, at $z=3$.
The five curves in each panel correspond
to $10\%$, $25\%$, $50\%$, $75\%$, $90\%$ percentiles;
i.e., $10\%$ of clouds has a temperature below the bottom curve,
while $90\%$ of clouds has a temperature below the top curve, etc.
\label{fig4}}
\end{figure*}

To further demonstrate that GSW do not significantly
alter the flux distribution of the \lya forest,
Figure 3 shows the probability distributions of transmitted flux fraction,
defined as $F\equiv \exp(-\tau)$,
for the seven runs tabulated in Table 1.
The fact that all the runs, except Runs 5 (N864L22M) and 7 (N432L11M32),
nearly overlay with one another
clearly shows that the effect of GSW on flux distribution
and other derived quantities (such as column density distribution, etc.,)
will remain relatively unaltered,
retaining the previous good agreement found between simulations
and observations.
The fact that the higher resolution run (Run 4: N864L11M)
agrees with lower resolution runs (Runs 1,2,3: N432L11M, N432L11L, N432L11H)
suggests that our fiducial run (Run 1: N432L11M) has adequate
resolution.
The deviation of Run 5 (N864L22M) from the rest is due to cosmic variance,
while the deviation of Run 7 (N432L11M32) from the rest is a result of
missing small-scale power in that run.

However, the fact that GSW do propagate some distance,
especially into the low density regions, as shown in Figure 1,
suggests that some low column density \lya clouds should be
affected to varying degrees.
Figure 4 shows the temperature as a function of \lya cloud
column density for
the two cases with
(left panel; Run 1: N432L11M) and without (right panel; Run 2: N432L11L)
GSW, respectively.
It is evident that, while \lya clouds with
column density $N_{HI}\ge 10^{14}$cm$^{-2}$ are
only affected modestly,
those with $N_{HI}\le 10^{14}$cm$^{-2}$
are increasingly affected.
A closer examination suggests that roughly $25\%$ of
clouds with $N_{HI}\le 10^{14}$cm$^{-2}$
is seen to experience significant heating by the GSW,
and the effect decreases towards higher columns.
However, we have checked the cloud velocity 
width distributions with the and without
GSW and do not find noticeable differences,
suggesting an overall domination of peculiar velocities
in broadening clouds.
Combining observations of hydrogen lines with metal lines, 
which suffer less thermal boradening,
however, may allow one to see the GSW heating
effect in the low column density \lya forest,
at least for some individual cases with low peculiar
velocity broadening.

\begin{figure*}[t!]
\centering
\begin{picture}(500,200)
\psfig{figure=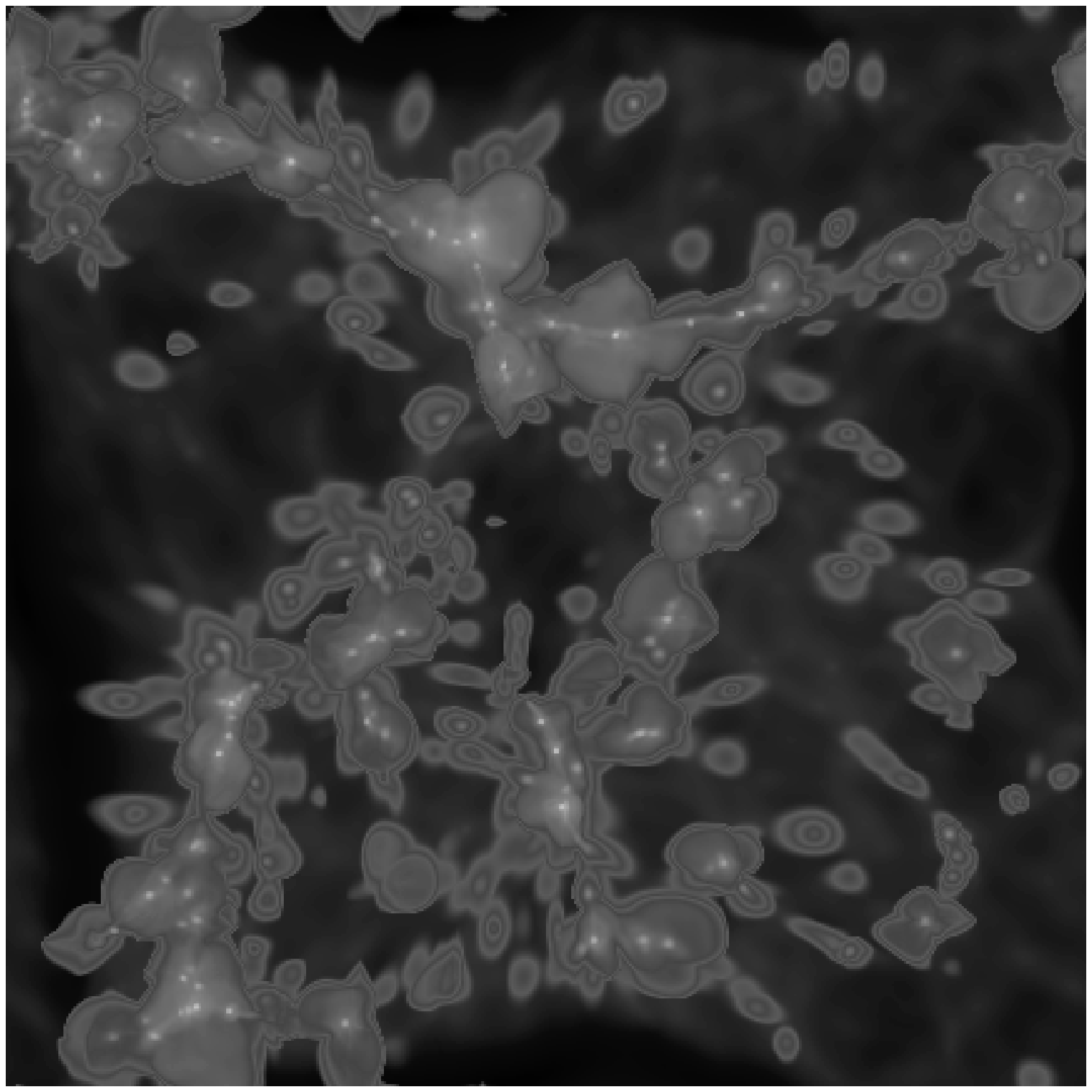,height=7.0cm,width=7.0cm,angle=0.0}
\end{picture}
\vskip -0.5cm
\centering
\begin{picture}(-50,-1000)
\psfig{figure=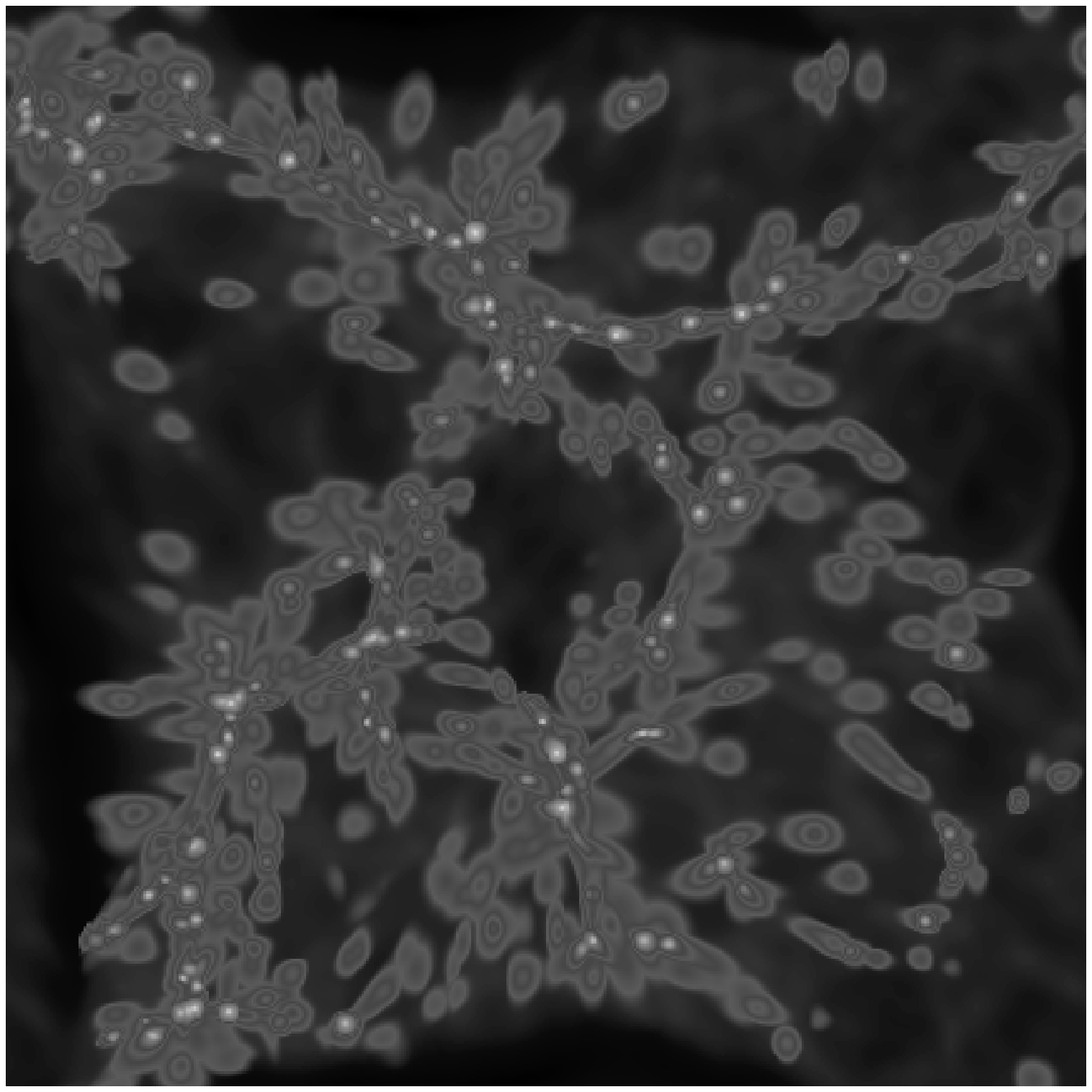,height=7.0cm,width=7.0cm,angle=0.0}
\end{picture}
\vskip -0.3cm
\centering
\begin{picture}(63,-1000)
\psfig{figure=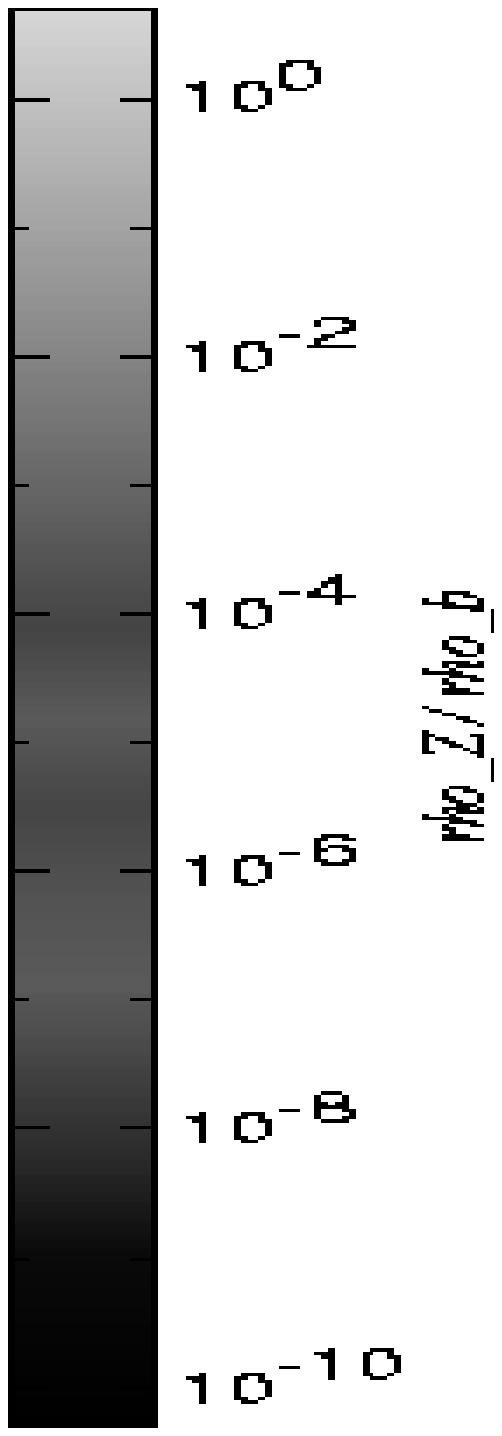,height=7.5cm,width=1.5cm,angle=0.0}
\end{picture}
\caption{
Projected metallicity of a slice of size $11\times 11h^{-2}$Mpc$^2$ comoving
and a depth of $2.75 h^{-1}$Mpc comoving at redshift $z=3$
for a {\it WMAP}-normalized $\Lambda$CDM model
with (left panel) and without (right) GSW, respectively.
The strength of the GSW is normalized to LBG observations.
This is the same slice as in Figure 1.
\label{fig5}}
\end{figure*}

\begin{figure*}[t!]
\centering
\begin{picture}(500,200)
\psfig{figure=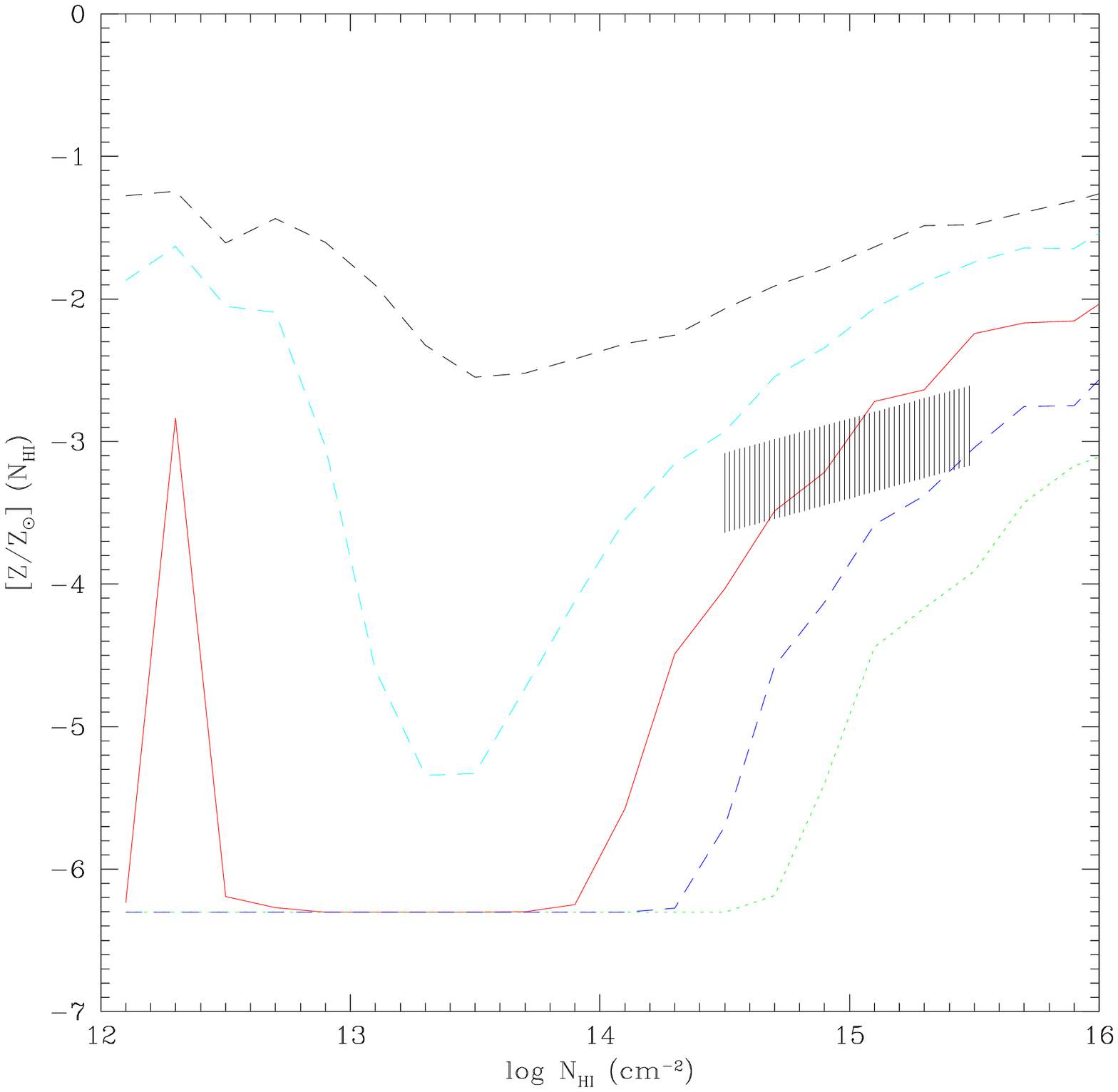,height=7.0cm,width=7.0cm,angle=0.0}
\end{picture}
\vskip -0.5cm
\centering
\begin{picture}(-50,-1000)
\psfig{figure=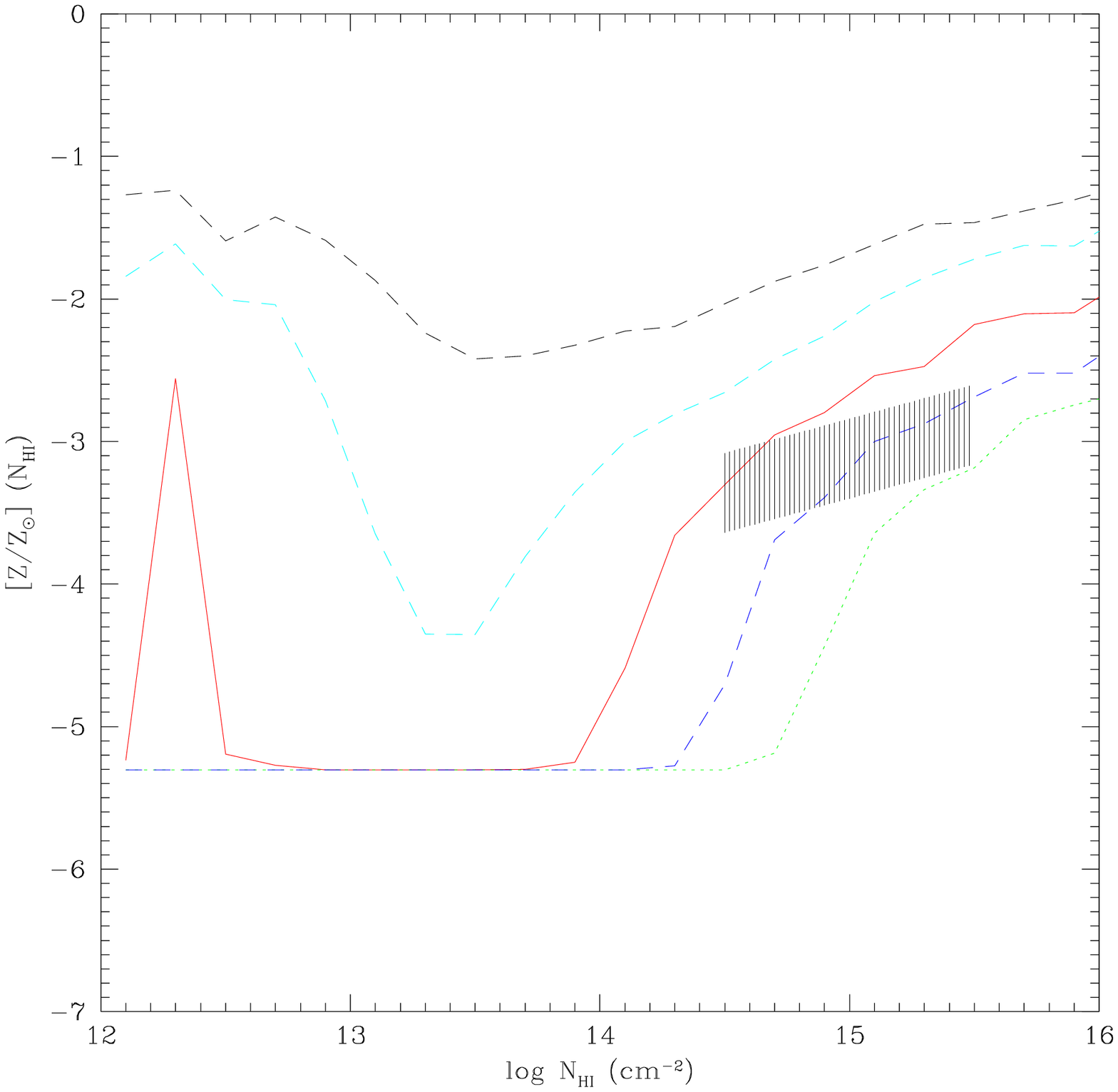,height=7.0cm,width=7.0cm,angle=0.0}
\end{picture}
\vskip -0.3cm
\centering
\caption{
shows the metallicity as a function of \lya cloud
column density for
Run 1 (N432L11M) with two cases  of metal yields:
left panel with constant yield $y_0=0.02$
and right panel with varying yield with a transition
to higher yield at $Z=10^{-3}\zsun$, at $z=3$.
The five curves in each panel correspond
to $10\%$, $25\%$, $50\%$, $75\%$, $90\%$ percentiles.
The shaded regions indicate the observed median metallicity
as a function of column density from Schaye \etal (2003)
with $1\sigma$ bounds.
\label{fig6}}
\end{figure*}

Let us now turn to the main point of the paper.
Could the GSW transport metal enriched gas to
raise the metallicity of low density regions to
a level consistent with the observed metallicity?
Are there palpable signatures of GSW on \lya forest?
Figure 5 shows the spatial distribution of metallicity in
the IGM with (left panel) and without (right panel) GSW.
It is visible
from Figures 5 that, while other, gravitational (e.g., Gnedin 1998) and
hydrodynamic processes do transport
metals to the vicinity ($\le \sim 100$kpc)
of galaxies without GSW (right panel of Figure 5),
GSW appear to play a more important role to transport
the metals from galaxies  to larger distances,
in conjunction with other, gravitational and non-gravitational
processes.
The ``metal bubbles" (reddish bubbles seen in the left panel of Figure 5)
have $\rho_{metals}/\rho_{gas}\sim 10^{-4}$, indicating
that these metal-contaminated regions are enriched to a metallicity close to
$10^{-2}\zsun$.

Figure 6 shows the metallicity as a function of \lya cloud
column density for our fiducial run (Run 1: N432L11M) with two
yield schemes.
For the clouds within the range of column densities
($N_{HI}\sim 10^{14}-10^{15} {\rm cm}^{-2}$),
where comparisons with observations
can be made,
it is very encouraging that
the agreement between observations and simulations
is good,
considering that our simulations have essentially only
one free parameter for the metal yield, which in turn
is fixed based on theory of stellar interior and
turns out also to be required to match the metallicity
of the intra-cluster gas
(Arnaud \etal 1994;
Mushotsky \etal 1996;
Mushotsky \& Lowenstein 1997;
Cen \& Ostriker 1999b).
A comparison between the left and right panels
suggests that, if there is a transition such that
the metal yield from stars is significantly higher for
nearly metal free gas, it seems that the transition
is likely to have occurred at $Z\le 10^{-3}\zsun$,
perhaps at $Z\sim 10^{-4}-10^{-3}\zsun$;
a transition at a higher gas metallicity would
over-enrich the IGM at the relevant densities.

\begin{figure*}[t!]
\centering
\begin{picture}(500,200)
\psfig{figure=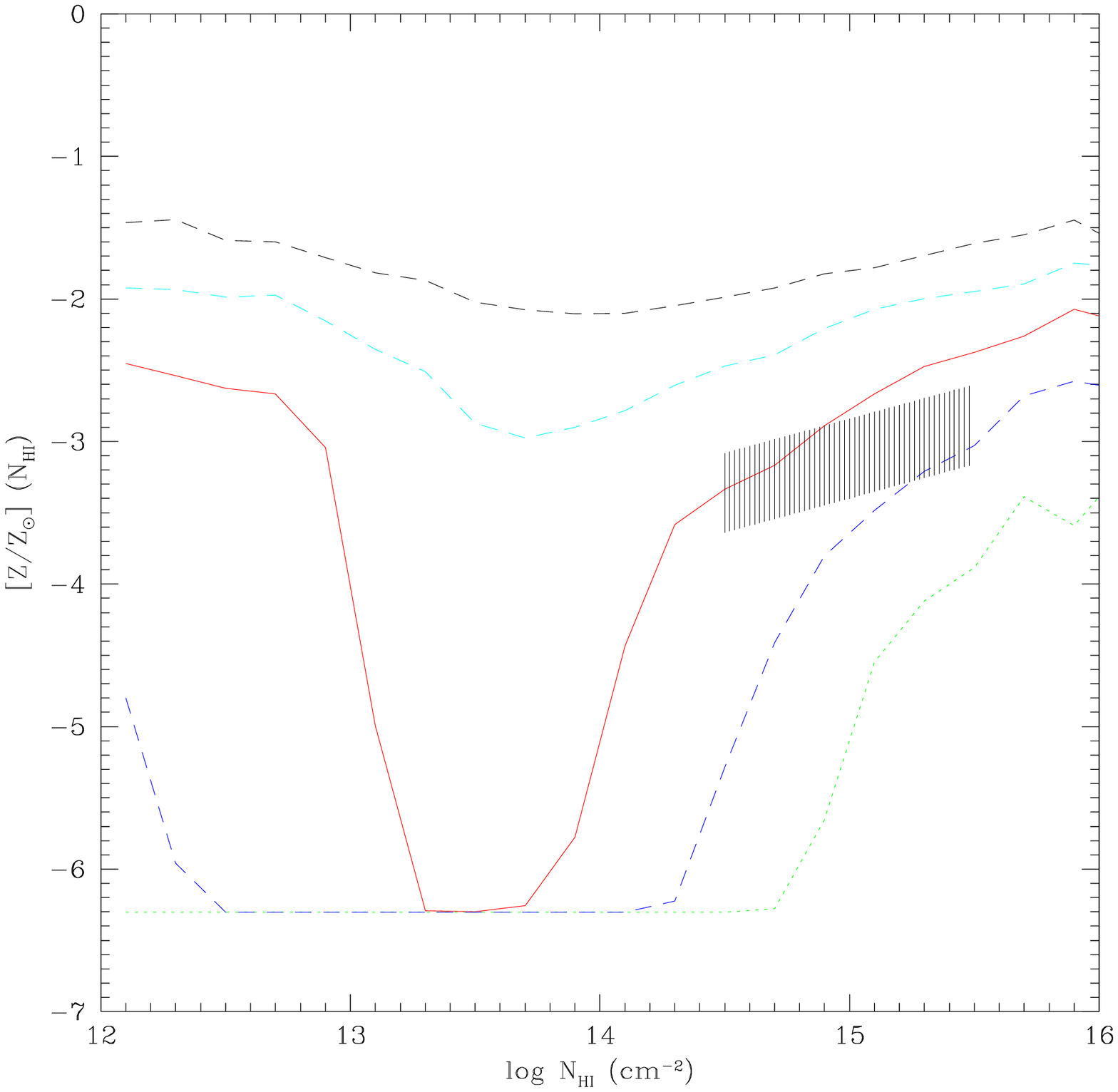,height=7.0cm,width=7.0cm,angle=0.0}
\end{picture}
\vskip -0.5cm
\centering
\begin{picture}(-50,-1000)
\psfig{figure=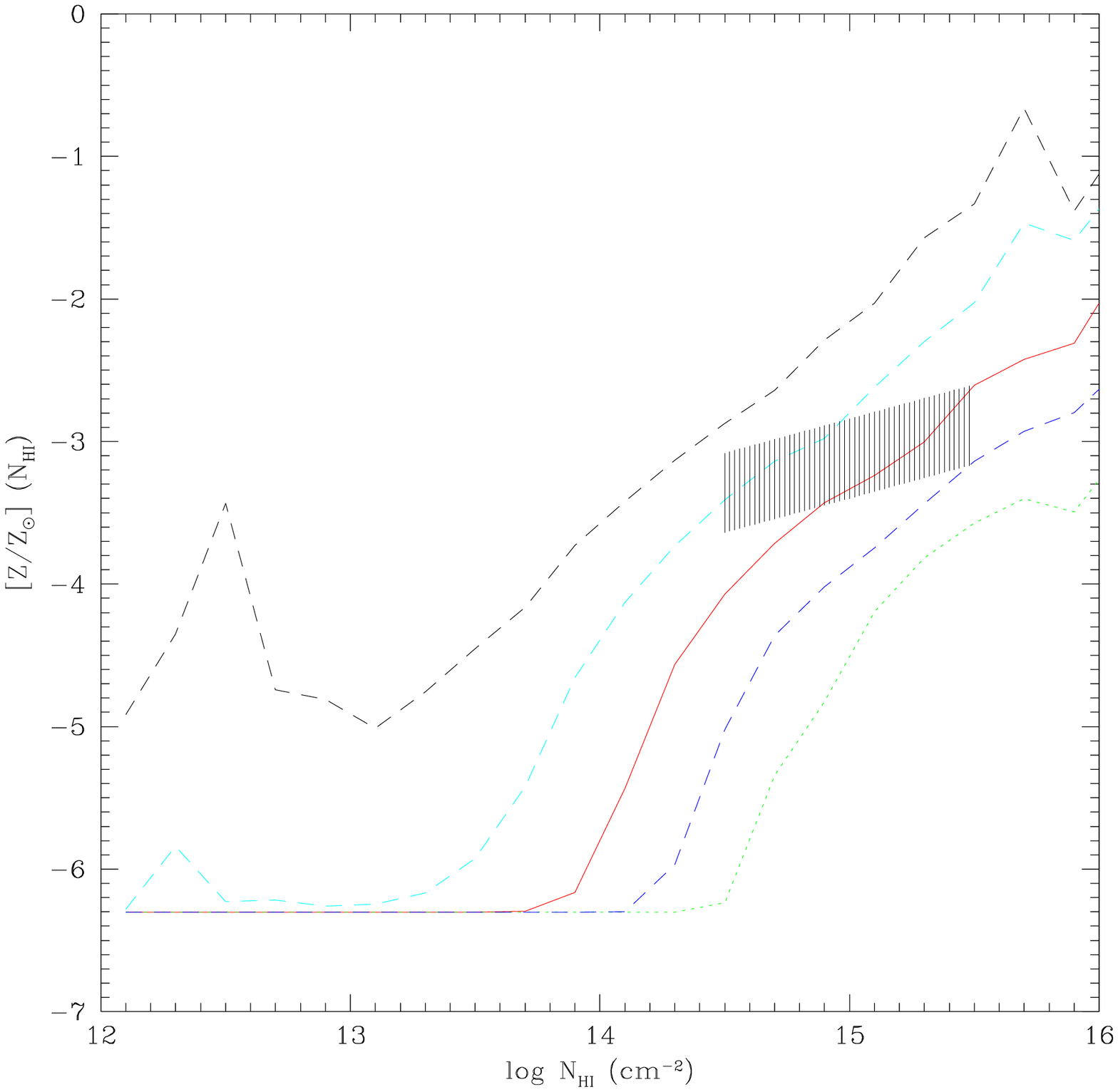,height=7.0cm,width=7.0cm,angle=0.0}
\end{picture}
\vskip -0.3cm
\centering
\caption{
shows the metallicity as a function of \lya cloud
column density for
Run 3 (N432L11H, left panel) and Run 2 (N423L11L, right panel) with constant
metal yield, at $z=3$.
The five curves in each panel correspond
to $10\%$, $25\%$, $50\%$, $75\%$, $90\%$ percentiles.
The shaded regions indicate the observed median metallicity
as a function of column density from Schaye \etal (2003)
with $1\sigma$ bounds.
\label{fig7}}
\end{figure*}

How sensitive are the results to the strength of GSW?
Figure 7 shows the cases with GSW strength (energy)
five times stronger than the fiducial run (left panel)
and with no GSW (right panel).
We see that both these cases are consistent with observations.
This is because the \lya clouds with column densities
in the range examined, where
metallicity can be observationally accurately determined,
are mostly located in the filaments somewhat further away from
galaxies and not substantially affected by GSW, consistent with Figure 3.
This indicates the metallicity of gas in \lya clouds with column 
density in the range $N_{HI}=10^{14.5}-10^{15.5}$cm$^{-2}$
mainly reflects the {\it local} star formation history.
Some effect of GSW is seen in the sense
that higher GSW produces somewhat higher metallicity
for the \lya clouds in that column density range,
but the differences are comparable to observational uncertainties. 
As it turns out, metallicity of \lya clouds in the column density
range of $N_{HI}=10^{14.5}-10^{15.5} {\rm cm}^{-2}$ provides an 
insensitive test of GSW (see Figure 8 for a further demonstration).

A more powerful discriminant may lie in the metallicity
of lower column density clouds,
to some of which GSW are able to
transport metals, as visually seen in Figures (1,5).
A closer comparison between left panels of Figures 6,7 (with GSW)
and right panel of Figure 7 (without GSW)
already reveals this signature:
{\it there are dramatic differences at $N_{HI}\le 10^{13.5} {\rm cm}^{-2}$,
between simulation with GSW and without GSW},
where $N_{HI}=10^{13.5} {\rm cm}^{-2}$ approximately
corresponds to $\rho/\langle\rho\rangle\le 1$ at $z=3$,
using the formula relating column density to gas density
in Schaye \etal (2003),
$\rho/\langle\rho\rangle = 10
(N_{HI}/10^{15}\hbox{cm}^{-2})^{2/3}[(1+z)/4]^{-3}$. Apparently GSW are able
to transport metals to some low density regions within which embedded star
formation was inefficient,
presumably in directions roughly perpendicular to the filaments
as seen in Figures (1,5).
In the fiducial run with GSW (Run 1: N432L11M) there are about $25\%$ of
\lya clouds with $N_{HI}<10^{13} {\rm cm}^{-2}$ may have metallicity
in excess of $10^{-2}\zsun$,
whereas there is none in the run without GSW.

\begin{figure*}[t!]
\centering
\vskip 0.7cm
\begin{picture}(250,200)
\psfig{figure=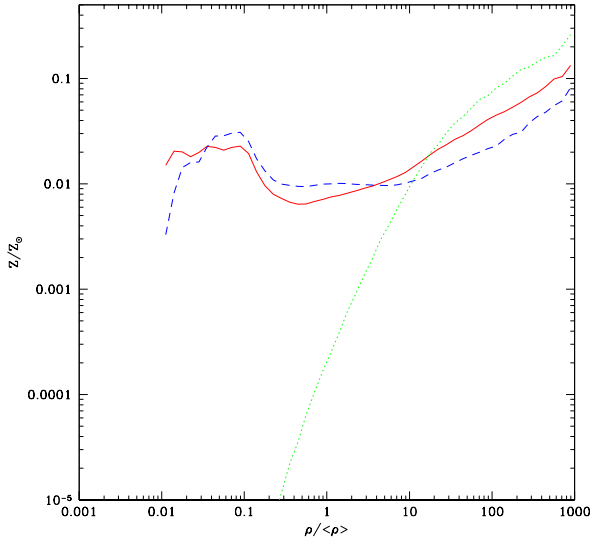,height=8.0cm,width=8.0cm,angle=0.0}
\end{picture}
\vskip -0.3cm
\centering
\caption{
shows mean metallicity as a function of gas density
for three runs: fiducial with realistic GSW (Run 1:N432L11M; solid curve),
high GSW run (Run 3:N432L11H; dashed curve)
and no GSW run (Run 2:N432L11L, dotted curve) at $z=3$.
\label{fig8}}
\end{figure*}

To elucidate this physical point, in Figure 8,
we plot the mean metallicity as a function of gas density.
We see the expected but now precisely quantified difference
between runs with GSW (solid and dashed curves)
and without GSW (dotted curve):
GSW are able to transport metals to regions
at $\rho/\langle\rho\rangle<10$,
whereas without GSW most of the metals are trapped
in regions with $\rho/\langle\rho\rangle>10$.
The mean metallicity is larger
by a factor of
$(7,40,500)$ at $\rho/\langle\rho\rangle=(3, 1, 0.1)$ in Run 1 (N432L11M)
than in Run 2 (N432L11L), with the difference becoming still larger at lower
$\rho / \langle\rho\rangle<0.1$. We see that without GSW mean metallicity is a
steep monotonic function of density,
whereas with GSW there two are peaks,
where the lower metallicity peak at $\rho=(0.01-0.1)\langle\rho\rangle$
represents metal enriched low density regions,
a signature of GSW, consistent with Figures (6,7).
This is the most clear demonstration of the GSW effect
on the metal enrichment of the IGM and a clear signature of GSW:
with increasing observational sensitivity one
should expect to see the metallicity at lower density regions
$\rho / \langle\rho\rangle < 1$  increasing rather than decreasing.
The existence of a metallicity trough at $N_{HI}=10^{13-14} {\rm cm}^{-2}$ 
(Figures 6, 7) or $\rho / \langle\rho\rangle=0.1-1$ (Figure 8)
in the simulations with GSW is a clear indication
that GSW propagate anisotropically;
in other words, some intermediate density region along filaments
are relatively less affected.
We note that metal enrichment from galaxies not resolved in our
simulations (a few times $10^7\msun$) at earlier epochs
is unlikely to be important compared to the observed levels,
as shown by Norman, O'Shea, \& Paschos (2004),
but may put a somewhat higher metallicity floor for the case without GSW.

\begin{figure*}[t!]
\centering
\vskip 0.7cm
\begin{picture}(250,200)
\psfig{figure=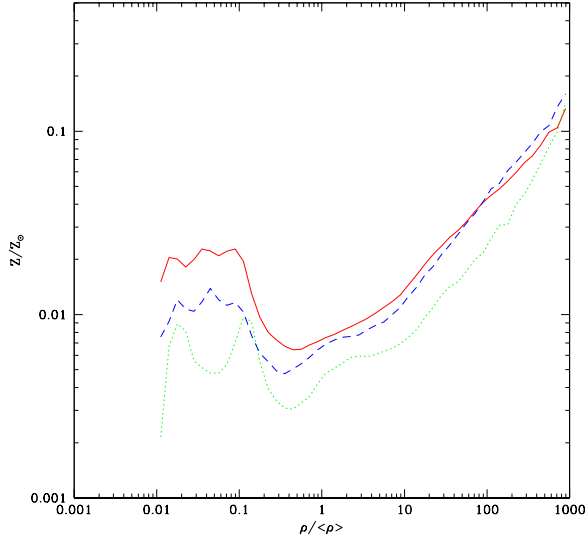,height=8.0cm,width=8.0cm,angle=0.0}
\end{picture}
\vskip -0.3cm
\centering
\caption{
shows mean metallicity as a function of gas density
for three runs: fiducial run with realistic GSW (Run 1:N432L11M; solid
curve), reduced small-scale power run at 8 cells (Run 6:N432L11M8; dashed
curve) and
reduced small-scale power run at 32 cells (Run 7:N432L11M32; dotted curve).
\label{fig9}}
\end{figure*}

What galaxies are responsible for transporting metals to
the low density regions?
Figure 9 shows a comparison between Runs 1,6,7 (N432L11M, N432L11M8,
N432L11M32). These vary in their high wavenumber cutoff corresponding
to minimum halo mass $M_{min}\sim (\pi/k_{max})^3\langle\rho\rangle$
of $1.1\times 10^7h^{-1}\msun$,
$7.0\times 10^8h^{-1}\msun$,
$4.5\times 10^{10}h^{-1}\msun$,
in Run 1 (N432L11M), 6 (N432L11M8) and 7 (N432L11M32), respectively. We see
that,
while the difference at $\rho/\langle\rho\rangle\sim 10^3$ where large
galaxies are located is small between the runs,
the difference between Run 1 (N432L11M) and Run 7 (N432L11M32)
is about $2.5-5.0$
and the difference between Run 1 (N432L11M) and Run 6 (N432L11M8) is about
$2$ at the low metallicity peak at $\rho/\langle\rho\rangle\sim 0.01-0.1$. A
simple interpretation of Figure 9 is
that galaxies with mass in the range
$(1.1\times 10^7h^{-1}-4.5\times 10^{10})h^{-1}\msun$
all contribute to the metal enrichment of the lower density IGM,
with approximately $25\%$ from
$>4.5\times 10^{10}h^{-1}\msun$,
$50\%$ from
$7.0\times 10^8-4.5\times 10^{10}h^{-1}\msun$
and
$25\%$ from $<7.0\times 10^{8}h^{-1}\msun$.
The recent work by Aguirre \etal (2001)
has shown that massive galaxies at low redshift are not very
effective in enriching the IGM to a relatively
uniform degree.
On the other hand,
GSW from dwarf galaxies at high redshift
appear to be able to more effectively disperse metals
relatively uniformly without traveling a very long distance
(e.g., Schwarz, Ostriker, \& Yahil 1975;
Cen \& Bryan 2001; Madau, Ferrara, \& Rees 2001).
Our results are fully consistent with these earlier works.
We further analyze the simulations by removing
a sphere of radius $1h^{-1}$Mpc
around each simulated Lyman Break Galaxy,
identified as galaxies brighter than rest-frame $V$-band
magnitude $M_V=-21$, as most of the brightest galaxies in the
simulation satisfy the color-color selection criteria of LBGs used by
observers (e.g. Nagamine 2002; Nagamine \etal 2004a,b).
The results (not shown) in a similar plot to Figure 6
are virtually identical to Figure 6.
This indicates that the contribution from ongoing
star forming massive galaxies is, as expected,
negligible, simply because
there is a lag due to finite GSW propagation time.
This is also in part because the massive galaxies
do not make large contribution to metal enrichment of the low density IGM,
consistent with Figure 9 and earlier results of Aguirre \etal (2001).

To better understand responsible galaxies for the metal enrichment  of \lya
forest, Figure 10 shows the ratio of mean secondary metallicity 
to mean primary metallicity as a function of column density.
Within galaxies, the ratio of secondary to primary 
metals is proportional to the ratio primary/hydrogen.
Thus higher values of S/P indicate 
an origin of metals in more massive, more metal rich systems. 
The most striking feature in this figure is the
dramatic decrease of the ratio from $\sim 1$ in the highest
column density clouds to about $0.003-0.02$ in the low
column density clouds, a drop of a factor of $50-300$,
for the fiducial model (red curve).
Quantitatively, we see that the ratio of
secondary (e.g., N) to primary metals (e.g., O,C)
is expected to be smaller by a factor of $10$
in clouds of $N_{HI}\sim 10^{14.5} {\rm cm}^{-2}$ compared to that
in large galaxies and by a factor of $\ge 50$ for 
$N_{HI}\le 10^{13.5} {\rm cm}^{-2}$.
This can be most easily understood and consistent with Figure 9,
if most of the metal enrichment of \lya forest
is due to dwarf galaxies (Dekel \& Silk 1986;
Mac Low \& Ferrara 1999; Madau, Ferrara, \& Rees 2001),
where gas retainment is more difficult
and thus metal recycling is limited.

\begin{figure*}[t!]
\centering
\vskip 0.7cm
\begin{picture}(250,200)
\psfig{figure=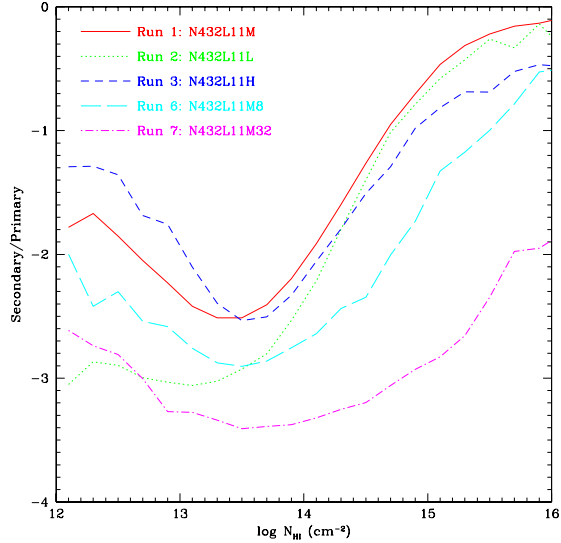,height=8.0cm,width=8.0cm,angle=0.0}
\end{picture}
\vskip -0.3cm
\centering
\caption{
shows secondary to primary metal ratio as a function of column density for
five runs at $z=3$.
The units on the y-axis is relative and normalized to have the maximum of
unity (normally in the densest regions).
\label{fig10}}
\end{figure*}

Some other interesting features are also present in Figure 10.
A comparison among the red solid, blue long-dashed
(with GSW) and green dotted curves (no GSW)
reveals a couple of noticeable properties.
First, the runs with GSW show an upturn of the ratio
at $N_{HI}<10^{13.5} {\rm cm}^{-2}$, although the level is still a factor  of
$50-100$ below the high density regions.
The fact there is an upturn and a valley at
$N_{HI}\sim 10^{13.5} {\rm cm}^{-2}$ shows that
the metals transported to the lowest density regions with
$N_{HI}<10^{13.5}$cm$^{-2}$ are somewhat more recycled through stars than
regions at $N_{HI}\sim 10^{13.5} {\rm cm}^{-2}$ 
and originate in higher metallicity systems.
This may be explained if the metals are largely transported by winds from
galaxies at densest peaks, where relatively more recycling has occurred,
whereas  near $N_{HI}\sim 10^{13.5} {\rm cm}^{-2}$ star formation and enrichment
are largely local and recent.
Second, increasing the strength of GSW causes more-recycled metals to be
transported to low density regions, as expected.
Comparison of the two runs with reduced small-scale power
(Runs 6,7; long-dashed cyan and dot-dashed magenta curves)
and the fiducial run (Run 1, solid red curve)
shows that artificial removal of small-scale power thus low mass galaxies
significantly reduces the overall values of the ratio of secondary to
primary metals of \lya forest concerned here, while the upturn at the low
column density end is preserved. This is of course easily understandable,
since much of previous generation of stars can no longer form without the
small-scale power.
This again highlights the need to have high enough $k_{max}$ in the initial
density field.

\begin{figure*}[t!]
\centering
\begin{picture}(500,200)
\psfig{figure=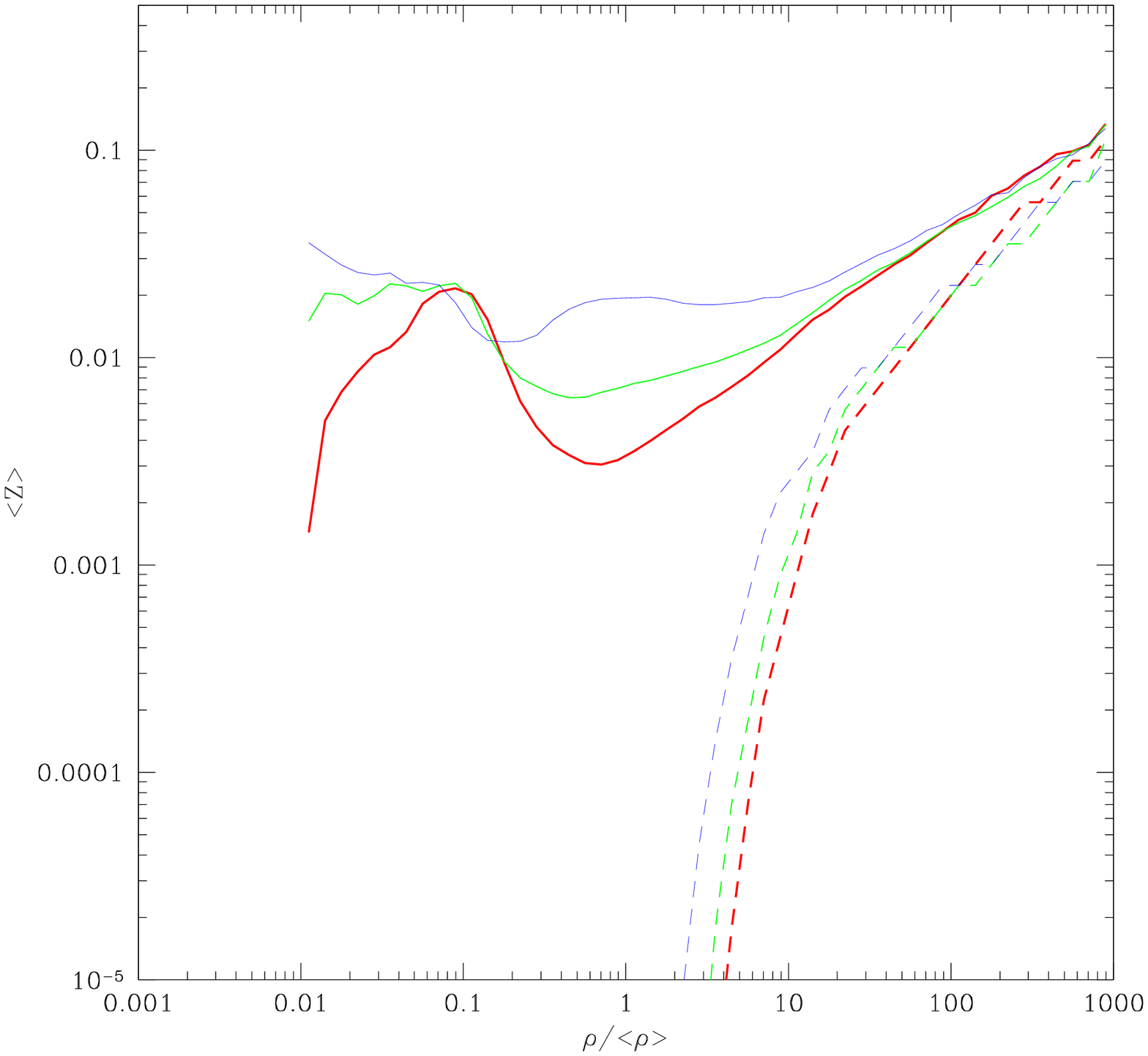,height=7.0cm,width=7.0cm,angle=0.0}
\end{picture}
\vskip -0.5cm
\centering
\begin{picture}(-50,-1000)
\psfig{figure=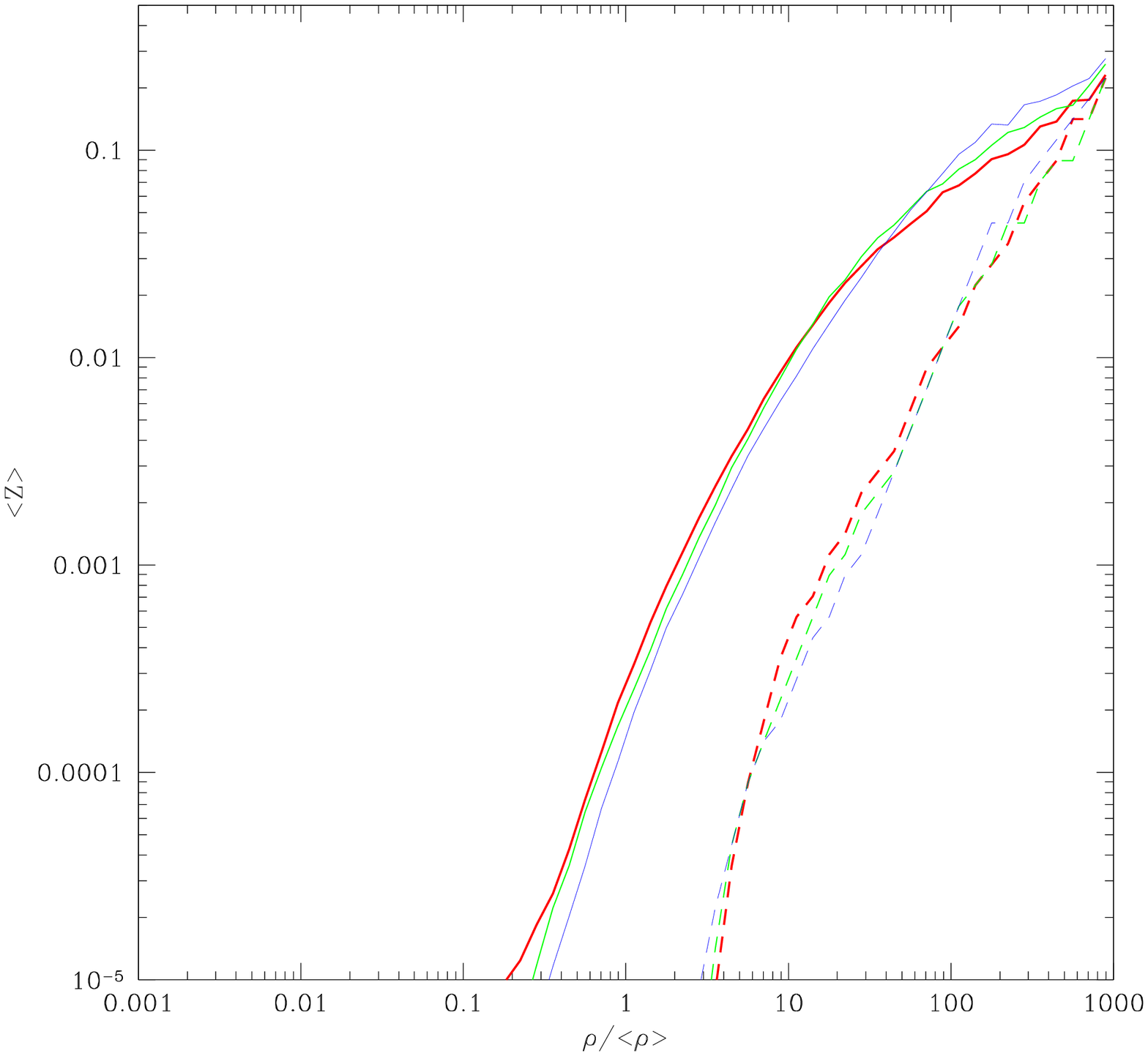,height=7.0cm,width=7.0cm,angle=0.0}
\end{picture}
\vskip -0.3cm
\centering
\caption{
shows the mean metallicity (solid curves)
and median metallicity (dashed curves)
as a function of \lya cloud
column density at redshift $z=2,3,4$ (thick to thin)
for Run 1 (N432L11M, left panel)
and for Run 2 (N432L11L, right panel).
\label{fig11}}
\end{figure*}

Finally, Figure 11 shows the mean and median metallicity
as a function of density at three redshifts,
$z=2,3,4$, with GSW (left panel)
and without GSW (right panel).
Both runs show only mild evolution in the median
metallicity, consistent with observations (Schaye \etal 2003):
most of the contamination of the IGM 
was completed at a relatively high redshift.
But the run with GSW shows considerable evolution
for the mean metallicity at $\rho/\langle\rho\rangle < 10$, whereas  the run
with GSW does not show significant evolution
even for the mean metallicity at all densities.

\section{Conclusions}

We use the latest high mass resolutions
hydrodynamic simulation of a $\Lambda$CDM model
to compute the metallicity evolution of the \lya forest.
Our primary goal is to investigate possible signatures of
galactic superwinds on the metallicity of the \lya forest.
There are three main points to be noted.

First, GSW do not significantly alter the flux distribution
of \lya forest and the agreement found in previous simulations
of cold dark matter model with observations remains unchanged.
On the other hand, GSW do increase the temperature of clouds
with column density $N_{HI}<10^{13.5} {\rm cm}^{-2}$, although
their contribution to the observed cloud width distribution
will be difficult to detect due to large peculiar velocities.

Second, the computed metallicity of \lya clouds in the column density range
of $N_{HI}\sim 10^{14.5}-10^{15.5} {\rm cm}^{-2}$ at $z=2-4$,
both with and without GSW,
is in reasonable agreement with observations (Schaye \etal 2003).
This suggests that these \lya clouds do not provide a sensitive
test of GSW.

Finally, we find a unique signature and sensitive test of GSW,
which lies in the still lower density regions
with gas density of $\rho/\langle\rho\rangle = 0.01-1.0$ and 
a corresponding column density of $N_{HI}\sim 10^{12}-10^{14} {\rm cm}^{-2}$.
Without GSW we predict that both the mean and median
metallicity of \lya clouds in this column density range at $z=2-4$ should
have $Z\le 10^{-3}\zsun$.
With GSW, however, there is a significant fraction ($\sim 25\%$)
of \lya clouds in this column density range which have a high metallicity
excess of $10^{-2}\zsun$, resulting in a mean metallicity
of $\sim 10^{-2}\zsun$.
If we (artificially) reduce the number of low mass galaxies
($M\le 4.5\times 10^{10}h^{-1}\msun$),
the contamination of the low column density clouds by GSW
is reduced by a factor of $\sim 4$, so it is likely
the mass and metal loss from these low mass systems
at $z>3$ (\cf Dekel \& Silk 1986)
is the origin of the metals.
There is a potential test of this hypothesis.
Since reprocessing of metals in these low mass systems
is negligible the ratio of secondary (e.g., N) to primary metals (e.g., O,C)
is very low and indeed, when we examine this tracer,
we find that the ratio of secondary to primary metals
is expected to be smaller by a factor of $10$ and $\ge 50$
for clouds of $N_{HI}\sim 10^{14.5} {\rm cm}^{-2}$ and 
$N_{HI}\le 10^{13.5} {\rm cm}^{-2}$, respectively, compared to that
in large galaxies.
Thus, future observations of N/O or N/C would help
provide an additional test of our proposal.
In addition, we find that there is a minimum in the median metallicity for
clouds of $N_{HI}\sim 10^{13}-10^{14} {\rm cm}^{-2}$ in the case with GSW,
whereas without GSW the metallicity decrease monotonically and rapidly with
decreasing column density.

\acknowledgments
We thank 
Len Cowie and Piero Madau 
for useful discussion.
This work is supported in part by grants AST-0206299 and NAG5-13381.

\newpage

\end{document}